\documentclass[reprint,prl,showpacs,superscriptaddress]{revtex4-2}
\usepackage{amssymb}
\usepackage{amsmath}    
\usepackage{graphicx}   
\usepackage{verbatim}   
\usepackage{gensymb}

\usepackage{xcolor}      
\usepackage{booktabs} 

\begin{document}
\title{Effects of colliding laser pulses 
polarization on $ e^{-}e^{+} $ cascade 
development in extreme focusing}

\author{M. Jirka}
\affiliation{Faculty of Nuclear Sciences and Physical Engineering, Czech Technical University in Prague, Brehova 7, 115 19 Prague, Czech Republic}
\affiliation{The Extreme Light Infrastructure ERIC, ELI Beamlines Facility, Za Radnici 835, 25241 Dolni Brezany, Czech Republic}

\author{S. V. Bulanov}
\affiliation{The Extreme Light Infrastructure ERIC, ELI Beamlines Facility, Za Radnici 835, 25241 Dolni Brezany, Czech Republic}
\affiliation{National Institutes for Quantum and Radiological Science and 
Technology (QST), Kansai Photon Science Institute, 8-1-7 Umemidai, Kizugawa, 
Kyoto 619–0215, Japan}

\begin{abstract}
	The onset and development of electron-positron cascade in a standing wave 
	formed by multiple colliding laser pulses requires tight focusing in order 
	to achieve the maximum laser intensity.
	There, steep spatio-temporal gradients in the laser intensity expel seed 
	particles from the high-intensity region and thus can prevent the onset of 
	a cascade.
	We show that radially polarized laser pulses ensure that the seed electrons 
	are present at the focal plane at the moment of the highest amplitude even 
	in the case of extreme focusing.
	This feature reduces the required laser power for the onset of a cascade 
	100 times (80 times) compared to circularly (linearly) polarized laser 
	pulses 
	having 
	the same focal spot radius and duration. 	
\end{abstract}

\maketitle

With the advent of multi-petawatt (PW) laser facilities, the possibility of 
creating electron-positron pairs via multiphoton Breit-Wheeler process in 
laser-electron collision has attracted a lot of attention 
\cite{Breit1934,Burke1997,Bamber1999,Blackburn2020,Gonoskov2022Charged}.
It is motivated by demand of studying the fundamental quantum electrodynamics 
(QED) processes present in nature using the extreme high intensity laser fields.
The great interest is especially getting to the regime of prolific 
electron-positron pair creation and energetic gamma flares generation, for 
which many possible experimental configurations have been proposed 
\cite{Ruffini2010,Blackburn2020,Gonoskov2022Charged}.
One widely accepted configuration for exploiting the radiation reaction effect 
and electron-positron pair production is the interaction of seed electrons with 
multiple laser beams \cite{Blackburn2020,Gonoskov2022Charged}.
In such a case, the electrons that are trapped in the standing wave (SW) formed by 
colliding laser beams can be re-accelerated after the recoil, and the photons 
emitted by nonlinear Compton scattering initiate the QED cascade \cite{Bula1996,Blackburn2020}
Provided the initially free seed electrons are trapped in the high intensity 
region, the cascade can develop at intensities $ \gtrsim 
10^{24}~\mathrm{W/cm^{2}} $ \cite{Bell2008} considering focal spot radius $w_0$ 
of several wavelengths $\lambda$ \cite{Tamburini2017}.
It corresponds to the laser power of hundreds PW.
However, today's multi-PW facilities provide the intensity of the order $10^{23}~\mathrm{W/cm^2}$ \cite{Danson2019,Yoon2021}.
Thus, laser intensities required for prolific electron-positron pair production can only be reached by compressing the laser energy into a tightly focused ($w_0\approx\lambda$) short laser pulse, i.e. in the so-called $\lambda^3$ interaction regime \cite{Mourou2002,Mourou2006}.
The natural consequence of tight focusing is that the steep spatio-temporal gradients in intensity can ponderomotively expel seed electron from the region of highest intensity and thus significantly suppress or even prevent the onset of radiation reaction and pair production \cite{Tamburini2017}.
As a result for $\lambda^3$ interaction volume, QED cascades are prevented even 
at intensities around $  10^{26}~\mathrm{W/cm^{2}} $ with tightly focused, 
linearly polarized laser pulses \cite{Tamburini2017}.
Therefore, the question of how to overcome the ponderomotive expulsion and keep 
the seed electron in the tiny region of the highest laser intensity in order to 
initiate the creation of electron-positron pairs is of paramount importance in 
strong ﬁeld QED.

The threshold for QED cascade can be lowered by the combination of multiple laser pulses and structured plasma targets \cite{Bulanov2010Multiple,Nerush2011,Gelfer2015,Grismayer2016,Vranic2016,Gong2017,Grismayer2017,Jirka2017}.
Polarization of colliding laser pulses affects the possibility of reaching the 
Schwinger limit field in the case, that the seeding electron is generated by 
the Schwinger mechanism in the focus of the SW 
\cite{Fedotov2010,Bulanov2010}.
The Schwinger (Sauter) limit field
$
E_{\mathrm{S}}=m_{e}^{2}c^{3}/e\hbar\approx1.33\times10^{16}~\mathrm{V/cm} 
$ represents the electric field strength needed for the creation of electron-positron pairs out of the vacuum \cite{Schwinger1951}.
Here $ m_{e} $ denotes the 
electron mass, $ c $ speed of light, $ e $ is the elementary charge 
and $ \hbar $ is the reduced Planck constant.
We note that the optimal focussing is achieved in a dipole field \cite{Bassett1986,Gonoskov2012,Gonoskov2013,Gonoskov2017,Efimenko2018,Magnusson2019,Efimenko2022,Marklund2023}, however, the QED cascade seeding requires a specific shape of a solid-density target \cite{Efimenko2019}.
Nevertheless, due to the debris damaging the optical elements, it is more reasonable to use a gas-jet target or initiate the QED cascade by stray electrons in imperfect vacuum \cite{Efimenko2019}.
In this respect, the mentioned configurations are not suitable since if the seed electron is placed in the center of the focal plane, i.e. where the highest intensity will be achieved, then the oscillating electric field of colliding laser pulses causes its expulsion away from the interaction region before the standing-wave structure is established.
Moreover, there exist significant differences in the dynamics of electron-positron pair production in a solid target and in a tenuous gas since the collective plasma effects in a solid target noticeably affect the electron dynamics and the QED cascade formation, whereas in a tenuous gas the dynamics is fully dominated by the laser fields \cite{Sampath2018}.
Hence, overcoming the ponderomotive expulsion of seed electrons from a tightly 
focused 
laser field, and thus lowering the threshold for the onset of QED cascade, 
remains the principal challenge.

\begin{figure}
	\centering
	\includegraphics[width=1.0\linewidth]{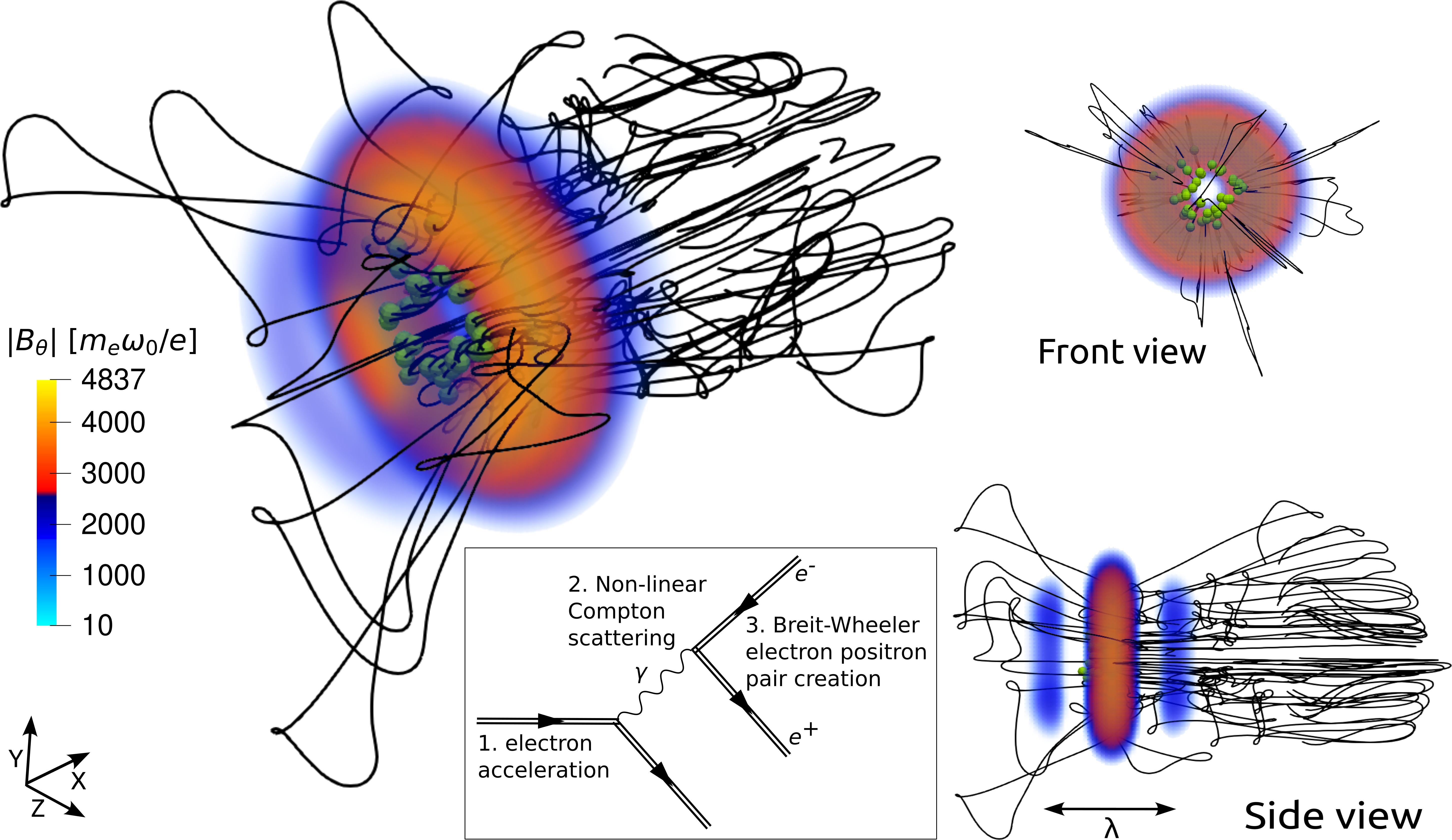}
	\caption{Position of seed electrons (green bullets) at the moment of the 
	highest laser intensity achieved in a collision of two counter-propagating 
	radially polarized laser pulses of intensity 
	$8.5\times10^{24}~\mathrm{W/cm^2}$. Solid lines show their preceding 
	trajectory. Inset: process diagram, double fermion lines 
	indicate that the process occurs in an external field.}
	\label{fig:01}
\end{figure}

In this paper, we study the onset of the QED cascade via the multiphoton 
Breit-Wheeler process, seeded by tenuous gas electrons, during a collision of 
two tightly focused laser pulses in the $ \lambda^3 $ regime.
It is shown that using radially polarized (RP) laser pulses reduces the 
required laser power for the onset of QED cascade by a factor of 80-100x 
compared to linear (LP) or circular (CP) polarization of the laser pulses 
having the same duration and focal spot radius $w_0$, respectively.
%
%
The uniqueness of the presented approach is that the colliding RP laser beams provide a strong restoring force that is acting in the opposite direction to the one in which the seed electrons have been previously ejected.
Therefore this feature ensures the cascade seeding even if the seed electrons were previously expelled from the center of the focal plane.
This anomalous phenomenon is not present in LP, CP or dipole 
\cite{Lindlein2007,Gonoskov2012,Gonoskov2013,Gonoskov2017,Efimenko2019,Bashinov2022}
 standing wave-like structures, and thus the seed electrons 
previously expelled from the center of the focal plane are not dragged back.
%
%
In addition, the proposed configuration represents a novel approach for obtaining relativistic positrons, as it enables to generate narrow multi-GeV bunches of positrons and electrons that are spatially separated and propagate in opposite directions.

The RP laser beam has radial and axial electric field components, while the magnetic field component is only azimuthal.
Such a spatial structure of the laser beam is useful especially for electron acceleration \cite{Salamin2002}.
Axially injected electrons are accelerated by the axial electric field component and confined by the azimuthally polarized magnetic field \cite{Salamin2006}.
Moreover, since the radial electric field component vanishes along the beam axis, the electrons are not diffracted out of the electron beam \cite{Salamin2006Electron}.
The RP laser pulse is also suitable for pair production in laser-electron collision as well as for reaching the range of parameters where a perturbative QED in strong external electromagnetic (EM) field breaks \cite{Jirka2022,Jirka2023Pair}

The EM field of a tightly focused fundamental RP laser pulse propagating along 
the $x$-axis is described by \cite{Salamin2007}
\begin{widetext}
	\begin{equation}\label{Br}
		E_{r}\left(x,y,z,t \right) =E_{0}e^{-r^{2}/w^{2}} \left[ \epsilon\rho 
		C_{2}+\epsilon^{3}\left(-\dfrac{\rho 
			C_{3}}{2}+\rho^{3}C_{4}-\dfrac{\rho^{5}C_{5}}{4} 
		\right)+\epsilon^{5}\left(-\dfrac{3\rho 
			C_{4}}{8}-\dfrac{3\rho^{3}C_{5}}{8}+\dfrac{17\rho^{5}C_{6}}{16}-\dfrac{3\rho^{7}C_{7}}{8}+\dfrac{\rho^{9}C_{8}}{32}
		\right)  \right],
	\end{equation}
	
	\begin{equation}\label{Bx}
		E_{x}\left(x,y,z,t \right)=E_{0}e^{-r^{2}/w^{2}}\left[\epsilon^{2}\left( 
		S_{2}-\rho^{2}S_{3}\right) + \epsilon^{4}\left( 
		\dfrac{S_{3}}{2}+\dfrac{\rho^{2}S_{4}}{2}-\dfrac{5\rho^{4}S_{5}}{4}+\dfrac{\rho^{6}S_{6}}{4}\right)
		\right],
	\end{equation}
	
	\begin{equation}\label{Etheta}
		B_{\theta}\left(x,y,z,t \right)=E_{0}e^{-r^{2}/w^{2}}\left[\epsilon\rho 
		C_{2}+\epsilon^{3}\left(\dfrac{\rho 
			C_{3}}{2}+\dfrac{\rho^{3}C_{4}}{2}-\dfrac{\rho^{5}C_{5}}{4} \right) 
		+\epsilon^{5}\left( \dfrac{3\rho C_{4}}{8} + \dfrac{3\rho^{3}C_{5}}{8} + 
		\dfrac{3\rho^{5}C_{6}}{16} - 
		\dfrac{\rho^{7}C_{7}}{4}+\dfrac{\rho^{9}C_{8}}{32}\right) 
		\right].
	\end{equation}
\end{widetext}
where for $ n=2,3, ... $
\begin{equation}
	C_{n}=\left( \dfrac{w_{0}}{w}\right) ^{n}\cos\left( \varphi+n\varphi_{G} 
	\right),
\end{equation}
and
\begin{equation}
	S_{n}=\left( \dfrac{w_{0}}{w}\right) ^{n}\sin\left( \varphi+n\varphi_{G} 
	\right).
\end{equation}
Here $ \epsilon=\lambda/\left( \pi w_{0}\right) $ is the diffraction factor \cite{Salamin2007}.
The radial distance from the beam axis $x$ is $ r=\sqrt{y^{2}+z^{2}} $, $
\rho=r/w_{0} $, $
w=w_{0}\sqrt{1+\left( x/x_{R}\right)^{2} } $, $ x_{R}=kw_{0}^{2}/2 $ is
the Rayleigh range, $ k=2\pi/\lambda $ is the wavenumber, $
\varphi=\varphi_{0}+\omega_{0} t - kx - kr^{2}/2R $, $
R=x+x_{R}^{2}/x $, $ \varphi_{0} $ is the initial phase and $
\varphi_{G}=\arctan\left( x/x_{R}\right)  $ is the Gouy phase.
For $ \varphi_{0}=0 $, the axial electric field $ E_{x} $ is zero in the
focal plane, while the radial electric $ E_{r} $ and azimuthal magnetic $
B_{\theta} $
fields reach their maximum here.
As a result, a doughnut-shaped transverse field structure is created in the focal
plane.
A quarter of a laser period later, the axial electric field $E_x$ reaches its 
maximum in the focal plane, while the radial electric field $E_r$ and azimuthal 
magnetic field $B_\theta$ vanish here.
The analytical expressions for EM fields of tightly focused LP laser pulse to order $ \epsilon^5 $ are given in Ref.~\cite{Salamin2007}. 
When RP laser pulses overlap, the dynamics of 
particles near laser axis ($x $ direction) is determined by $E_x 
\propto 
\cos(kx)\sin(\omega_{0}t)$ considering the initial phase shifts of two colliding 
lasers $ \varphi_{01}=0$ and $\varphi_{02}=\pi 
$ in Eqs.~\eqref{Br}-\eqref{Etheta}.
The field $ E_x $ is the strongest accelerating component and reaches its amplitude at antinodes along the $ x $ axis, i.e. when $ r=0 $.
The advantage lies in the direction of the field: at each two adjacent 
antinodes, $x = n\lambda/2 $, the field has opposite direction.
Thus the particle, that is expelled from one antinode, is consequently slowed 
down and accelerated backward at the nearby antinode.
On the contrary, in the case of LP, the particles along the laser pulse axis 
experience the field $E_y \propto \cos(kx)\sin(\omega_{0}t)$, that causes 
electron acceleration in a transverse direction that leads to electron ejection 
from the interaction region.
%

The role of ponderomotive force in the longitudinal direction is not crucial 
due to its opposite orientation at surrounding antinodes.
The ponderomotive force expells electrons also in transverse direction.
Thus, the accelerated electrons can reach high $ \chi_{e} $ values and emit photons when traverse the 
transverse fields $ B_\theta \propto \cos(kx)\cos(\omega_{0}t) $ and $ E_r 
\propto \sin(kx)\sin(\omega_{0}t) $ that are nonzero at radial distance $ r>0 $.
Since the transverse field has a doughnut-shaped structure, the ponderomotive 
force of transverse fields is pushing electrons back to the interaction region.
This unique and beneficial behaviour stems from convenient distribution of EM fields in RP 
SW structure and allows the feature of trapping particles in 
a high-intensity region as shown in Fig.~\ref{fig:01}.

In QED the interaction of electrons, positrons and photons with EM field is characterized by two Lorentz invariant parameters, $ \chi_{e,p}=\sqrt{-\left( F^{\mu\nu}p_{\nu}\right)^{2} }/m_{e}cE_{\mathrm{S}} $ ($\chi_{e}$ for electrons, $\chi_{p}$ for positrons) and $ \chi_\gamma= \hbar\sqrt{-\left( F^{\mu\nu}k_{\nu}\right)^{2}}/m_{e}cE_{\mathrm{S}}$ for photons, where $ F_{\mu\nu}=\partial_{\mu}A_{\nu}-\partial_{\nu}A_{\mu}$ is the EM field tensor, $ A_{\nu} $ is the four-potential, $ p_{\nu} $ is the four-momentum of the electron (or positron), and $ k_{\nu} $ is the photon four-wave vector \cite{Ritus1985}. 
If the photons emitted due to radiation reaction satisfy $\chi_{\gamma}\gtrsim 1$, non-linear Breit-Wheeler electron-positron pair creation becomes probable \cite{Breit1934,Gonoskov2022Charged}.

We studied the collision of two counter-propagating, tightly focused laser 
pulses having radial, linear and circular polarization in $ \lambda^3 $ regime 
using full-scale 3D particle-in-cell simulations in the code {\sf SMILEI} \cite{Derouillat2018}.
The tightly focused laser pulse is characterized by $ \lambda=1~\mathrm{\mu 
m}$,  $ w_0=\lambda/2 $ and gaussian temporal envelope of 
full-width-at-half-maximum duration $\tau= T/2 $ (in laser intensity), where 
$T=\lambda/c$ is the laser period.
%
%
In order to assess the efficiency of electron-positron pair production we keep 
the laser pulse energy, duration and $ w_0 $ fixed for all three cases of laser 
polarization at a given laser intensity in the range $ 
10^{24-26}~\mathrm{W/cm^{2}} $.
The seed particles are located in the center of the simulation box in a sphere of radius $ 0.02\lambda $ having a tenuous gas density $ 1\times10^{-6}n_{c}\approx1\times10^{15}~\mathrm{cm^{-3}} $, where $ n_c = m_e \varepsilon_0 \omega_0^2 / e^2  $ is the critical plasma density, $ \varepsilon_0 $ is the vacuum permittivity and $ \omega_0=2\pi c/\lambda $.

At the beginning of the interaction, the electrons, having zero initial momentum, are located in a small volume in the center of the focal plane.
When two tightly focused LP or CP laser pulses collide, these seed electrons 
are expelled in the radial direction by the ponderomotive force and thus do not 
experience the strong field region of the established SW.
On the contrary, colliding RP laser pulses ensure that the seed electrons are present at the focal plane at the moment of the highest amplitude.
Once the SW starts to be formed, the electrons located at the laser 
pulse axis are accelerated along the axial direction towards the center of one 
of the colliding laser pulses.
In the $\lambda^3$ regime of interaction, the electrons initially located in the focal plane at radial distances $ \lesssim \lambda/4 $ are predominantly accelerated along the $ x $-direction since here the axial field $E_x$ is stronger than the radial field $E_r$.
%
%
Thus the electron penetrates inside the counter-propagating laser pulse and since the axial electric field $E_x$ changes its sign every half-cycle and becomes stronger in the center of the laser pulse, the electrons are reversed and brought back to the focal spot region, where the SW structure reaches its amplitude.

To demonstrate this phenomenon, we show in Fig.~\ref{fig:01} the position of 
electrons (green bullets) in the SW at the moment when the maximum 
laser intensity is reached.
Black lines represent the preceding trajectories of these seed electrons. 
Before seed electrons reach the focal plane in the moment of the highest intensity of the established SW, they have been driven by the colliding laser pulses away from the focal spot in axial and also radial directions.
The asymmetry of particle trajectories along the axial $x$-direction (shown in 
a side view) is caused by the phase dependence of the axial field $E_x$.
Once the SW is established, the seed electrons in the focal plane 
experience the superposition of the azimuthal magnetic field components 
$B_\theta$, while the radial ones are canceled in the SW structure.
Thus seed electrons radiate photons that, in turn, produce electron-positron 
pairs.
We note that in multiple laser pulse configurations characterized by poloidal 
electric field and toroidal magnetic field structures in the focus, or vice 
versa, the axial component of the EM field quickly expels particles in the 
axial direction from the center of the focal spot 
\cite{Bulanov2010,Gonoskov2013,Bashinov2022}.	
Here we demonstrate the different field structure that is created by collision of two radially polarized laser pulses.
It is shown that in our case the axial components of colliding laser beams act in the opposite sense as they push seed electrons towards to a focal plane.
At the moment of the highest laser intensity, the axial field vanishes and electrons thus experience the amplitude of azimuthal magnetic field in the focal plane allowing a QED cascade to develop.

\begin{figure}
	\centering
	\includegraphics[width=1.0\linewidth]{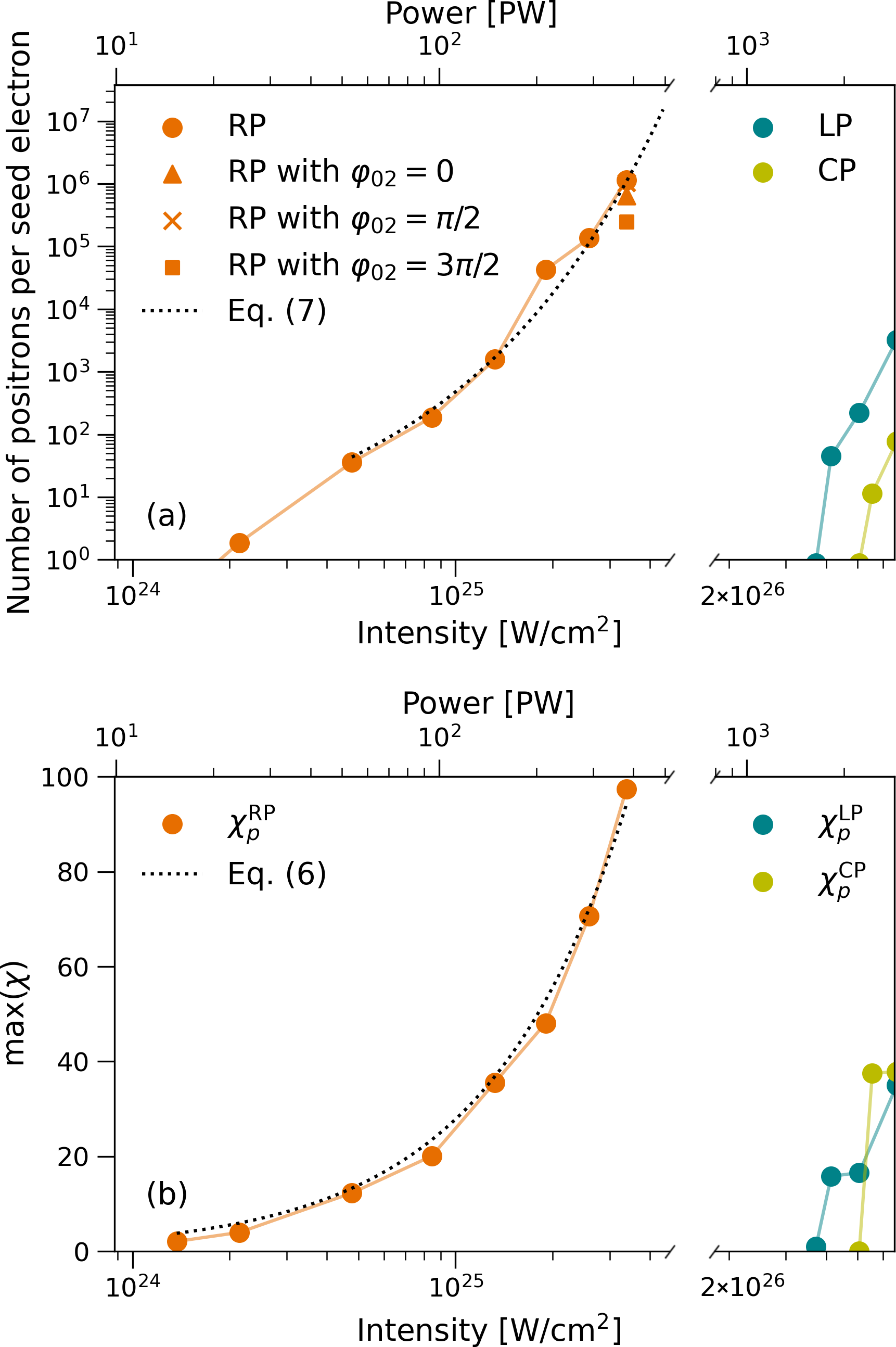}
	\caption{ (a) Number of 
	positrons 
	created in a collision of 
	two 
	counter-propagating RP, LP and CP laser pulses as a function of laser intensity (or power). Markers denote 3D results, dotted line represents the theoretical model.
	(b) Maximum values of $\chi_{p}$ parameter achieved by 
	created positrons for different laser pulse 
	polarization in 3D simulations (bullets) and according to theoretical model 
	(dotted line).}
	\label{fig:02}
\end{figure}

Figure~\ref{fig:02}(a) shows the final number of created positrons per one seed electron for RP, LP and CP according to 3D simulations.
Prolific production of positrons sets in at intensity $2\times10^{24}~\mathrm{W/cm^2}$ in the case of RP laser pulse which is two orders of magnitude below the threshold for pair production in the case of LP and CP laser pulse.
While the required laser power is 20~PW for RP laser pulse, the two latter polarizations require laser power that is 80-100x greater, respectively.
So far, we have assumed that the colliding laser pulses have the same phase and thus the SW is created by their perfect overlap.
To demonstrate the robustness of the interaction configuration with RP laser 
pulses, we consider different initial phase shift $\varphi_{02}$ of the second 
laser: $ 0 $, $ \pi/2 $ and $ 3\pi/2 $.
In these non-ideal cases, the field structure in the focal plane is given by the combination of all components of the radially polarized laser pulses.
In the worst case scenario $\varphi_{02}=3\pi/2$, the number 
of created 
positrons is lowered by a 
factor of five, as shown in Fig.~\ref{fig:02}(a), and thus the intensity threshold 
for the onset of
electron-positron pair creation is not significantly phase sensitive.
The theoretically expected intensity threshold $ I_{\mathrm{thr}} $ for 
electron-positron cascade for 
radially polarized laser pulses is $   
I_{\mathrm{thr}} = 2\times10^{24}~\mathrm{W/cm^{2}} $ 
($ E_{\mathrm{thr}} = 3.9\times10^{15}~\mathrm{V/m} $), which is in a good 
agreement with 3D simulation results shown in 
Fig.~\ref{fig:02}.
For derivation see Supplementary Material \cite{supp}.
For our setup, the maximum value of $\chi_{e,p}$ parameter that 
can be achieved by a particle is \cite{supp} 
\begin{equation}\label{max_chi}
\chi_{e,p}^{\mathrm{max}}\approx \dfrac{e \lambda I}{4 m_e c^2}\sqrt{\dfrac{2}{c\epsilon_0 
		I_S}},
\end{equation}
where $ I_S=2.32\times10^{29}~\mathrm{W/cm^{2}} $ denotes the Schwinger (Sauter) intensity.
This theoretical prediction is in excellent agreement with 3D simulation 
results presented in Fig.~\ref{fig:02}(b).
%
%
%
%
%
%
%
%
The number of created positrons is \cite{supp}
\begin{equation}\label{growth}
N_p = \exp\left[ (1.15/2) (E/E_\mathrm{thr}) \bar{\Gamma} 
\tau  \right].
\end{equation}
The growth rate is determined by 
$\bar{\Gamma}=\bar{W}_{p}/2(-1+\sqrt{1+8\bar{W}_{\gamma}/\bar{W}_{p}})$ \cite{Fedotov2010,Bashmakov2014,Grismayer2017}  where the probabilities of single photon emission $ \bar{W}_{\gamma} $  and pair production $ \bar{W}_{p} $ \cite{Mironov2021,Ritus1985} were calculated for the average values  $\bar{\chi}_{e,p} = 
\chi_{e,p}^{\mathrm{max}}/4$ and $\bar{\chi}_\gamma = 
\bar{\chi}_{e,p}/4$ and average energies $ 
\bar{\mathcal{E}}_e=\bar{\mathcal{E}}_\gamma=(e \lambda 
\sqrt{2I/c\varepsilon_0})/4 $ \cite{supp}.
%
%
%
As shown in Fig.~\ref{fig:02}(a), the 3D simulation results are in excellent 
agreement with the theoretical prediction given 
by 
Eq.~\eqref{growth} in the limit of its validity, i.e. $\chi_{e,p}\gg1$.
For $I> 10^{24}~\mathrm{W/cm^{2}} $, the growth rate of 
positrons satisfies the condition on the threshold of QED cascade development $ 
(1.15/2)(E/E_{\mathrm{thr}})\bar{\Gamma} t_{\mathrm{esc}} > 1 $ 
\cite{Fedotov2010,Grismayer2017}, where escaping time is $ t_{\mathrm{esc}} = 
w_0/c $.
This indicates the QED cascade development.
%
%
%
%

\begin{figure}
	\centering
	\includegraphics[width=1.0\linewidth]{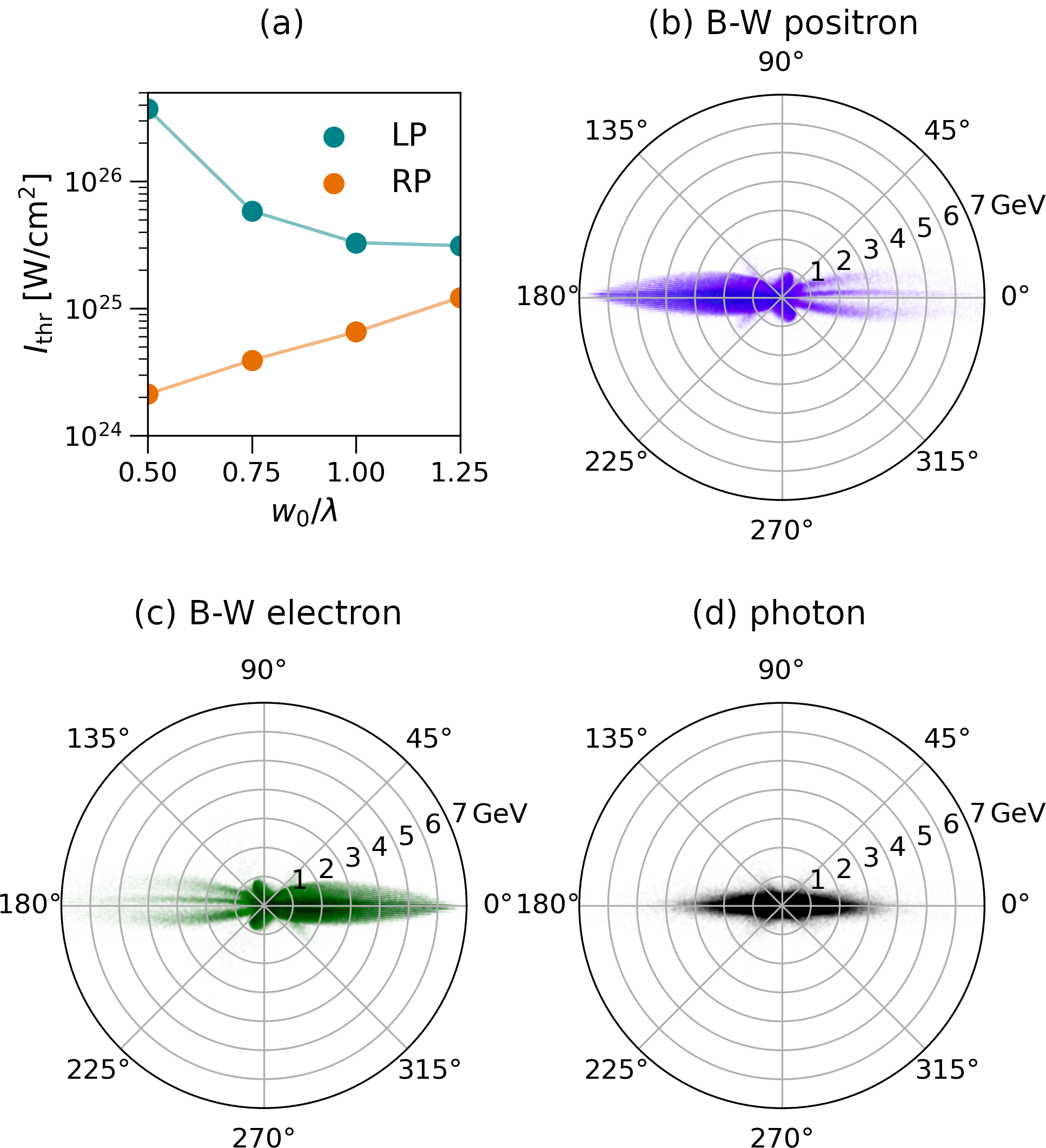}
	\caption{(a) Threshold intensity $ I_{\mathrm{thr}} $ for RP and LP as 
	obtained from 3D simulations for different values of a focal spot radius $ 
	w_0 $. Angular energy distribution of created (b) Breit-Wheeler
	positrons, (c) electrons and (d) photons in the $x,y$ 
	plane. The positive $x$ axis corresponds to the angle of $0\degree{}$. The 
	intensity of the radially polarized electric field component is 
	$1.9\times10^{25}~\mathrm{W/cm^2}$. }
	\label{fig:03}
\end{figure}

In Fig.~\ref{fig:03}(a) we show the dependence of threshold intensity $ 
I_\mathrm{thr} $ on the focal spot radius $ w_0 $ for RP and LP laser pulses as obtained from 3D simulations.
%
%
It demonstrates the advantage of RP for the generation of QED cascade by 
laser pulses focused beyond paraxial approximation.
In the case of $w_0=0.5\lambda$, the required power is 20~PW for RP, while for LP it is $\approx$1600~PW.
The required power for tightly focused RP is much lower than the power needed by widely focused RP or LP laser pulse (at $w_0=1.25\lambda$ corresponds to power $\approx$800 PW).

According to 3D simulations for $w_0=0.5\lambda$, even in the case that seed electrons initially occupy the sphere of a $\lambda$ diameter to fully cover the interaction region, the threshold for QED cascade in the case of RP is $2\times10^{24}~\mathrm{W/cm^{2}}$, while for LP it is $1\times10^{26}~\mathrm{W/cm^{2}}$.
This demonstrates the robustness of the proposed interaction with respect to the dimension of interaction volume.

We note that it is experimentally easier to generate LP laser beam compared to 
the RP one \cite{Salamin2008}.
Nowadays, the maximum intensity obtained with RP laser 
pulse is $ 10^{19}~\mathrm{W/cm^{2}} $ \cite{Carbajo2014}, and, therefore, 
further progress is needed to scale up intensities reachable with a 
radially polarized laser pulse, e.g. by using plasma-based optical phase 
modulators \cite{Leblanc2017,Porat2022} or by combining the beams 
\cite{Guo2023}. 
However, it has been experimentally demonstrated, that the 
radially polarized laser beam can be tightly focused to $ w_0=0.6\lambda $ 
reaching 
intensity $ 10^{17}~\mathrm{W/cm^{2}} $
\cite{Payeur2012}.

Due to the action of the strong axial electric field $ E_x $, the 
created 
electron-positron pairs are consequently accelerated to a multi-GeV level.
In Fig.~\ref{fig:03}(c,d) we show the angular distribution of created 
Breit-Wheeler positrons and electrons in the interaction of laser pulses whose 
radially polarized component has intensity $1.9\times10^{25}~\mathrm{W/cm^2}$.
When the laser pulses completely overlap, the axial electric field reaches the amplitude of $7.5\times 10^{25}~\mathrm{W/cm^2}$.
Assuming action of such a field for a period of $T/4$, 
electrons and positrons can be accelerated up to 6~GeV.
This is in agreement with 3D simulation results shown in Fig.~\ref{fig:03}.
The significant feature of the proposed interaction setup is that the newly created electrons and positrons are accelerated in narrow bunches propagating predominantly in mutually opposite directions.
Therefore, this configuration ensures generation of a separated bunch of 
relativistic positrons that can be consequently detected, as their 
number is orders of magnitude greater than the number of seed electrons.
As shown in Fig.~\ref{fig:03}(e), GeV photons are emitted in well-collimated 
flashes symmetrically along the axial direction.
The proposed scheme thus can serve also as a directed GeV photon source.

We show that properly chosen polarization of colliding laser pulses can provide 
the conditions for development and detecting of prolific electron-positron pair 
production in terrestial laboratory experiments.
According to 3D particle-in-cell simulations, the laser intensity required 
for generation of QED cascade seeded by a tenuous gas can be lowered by two 
orders of magnitude when radially polarized laser pulses are used instead of 
traditionally considered linearly or circularly ones.
%
%
It is given by the convenient orientation of the electric and magnetic fields of the colliding radially polarized laser pulses.
Even if the seed electrons initially located in the center of the focal plane 
are expelled away as the laser pulses start to form a SW, they are 
consequently dragged back to the focal plane just in the moment of reaching the 
highest intensity.
Moreover, the presented configuration generates narrow multi-GeV bunches of positrons and electrons that 
propagate in opposite directions and thus it facilitates positron detection.
Thus it can also serve as a directed GeV photon source offering new 
experimental possibilities in QED.

\begin{acknowledgments}
We would like to thank Prokopis Hadjisolomou for useful discussions.
Portions of this research were carried out at the ELI Beamlines
Facility, a European user facility operated by the Extreme Light
Infrastructure ERIC. 
Computational resources were provided by the e-INFRA CZ project (ID:90254), supported by the Ministry of Education, Youth and Sports of the Czech Republic.
\end{acknowledgments}
	
\bibliography{pair}{}

\begin{thebibliography}{53}%
\makeatletter
\providecommand \@ifxundefined [1]{%
 \@ifx{#1\undefined}
}%
\providecommand \@ifnum [1]{%
 \ifnum #1\expandafter \@firstoftwo
 \else \expandafter \@secondoftwo
 \fi
}%
\providecommand \@ifx [1]{%
 \ifx #1\expandafter \@firstoftwo
 \else \expandafter \@secondoftwo
 \fi
}%
\providecommand \natexlab [1]{#1}%
\providecommand \enquote  [1]{``#1''}%
\providecommand \bibnamefont  [1]{#1}%
\providecommand \bibfnamefont [1]{#1}%
\providecommand \citenamefont [1]{#1}%
\providecommand \href@noop [0]{\@secondoftwo}%
\providecommand \href [0]{\begingroup \@sanitize@url \@href}%
\providecommand \@href[1]{\@@startlink{#1}\@@href}%
\providecommand \@@href[1]{\endgroup#1\@@endlink}%
\providecommand \@sanitize@url [0]{\catcode `\\12\catcode `\$12\catcode
  `\&12\catcode `\#12\catcode `\^12\catcode `\_12\catcode `\%12\relax}%
\providecommand \@@startlink[1]{}%
\providecommand \@@endlink[0]{}%
\providecommand \url  [0]{\begingroup\@sanitize@url \@url }%
\providecommand \@url [1]{\endgroup\@href {#1}{\urlprefix }}%
\providecommand \urlprefix  [0]{URL }%
\providecommand \Eprint [0]{\href }%
\providecommand \doibase [0]{https://doi.org/}%
\providecommand \selectlanguage [0]{\@gobble}%
\providecommand \bibinfo  [0]{\@secondoftwo}%
\providecommand \bibfield  [0]{\@secondoftwo}%
\providecommand \translation [1]{[#1]}%
\providecommand \BibitemOpen [0]{}%
\providecommand \bibitemStop [0]{}%
\providecommand \bibitemNoStop [0]{.\EOS\space}%
\providecommand \EOS [0]{\spacefactor3000\relax}%
\providecommand \BibitemShut  [1]{\csname bibitem#1\endcsname}%
\let\auto@bib@innerbib\@empty
\bibitem [{\citenamefont {Breit}\ and\ \citenamefont
  {Wheeler}(1934)}]{Breit1934}%
  \BibitemOpen
  \bibfield  {author} {\bibinfo {author} {\bibfnamefont {G.}~\bibnamefont
  {Breit}}\ and\ \bibinfo {author} {\bibfnamefont {J.~A.}\ \bibnamefont
  {Wheeler}},\ }\href {https://doi.org/10.1103/physrev.46.1087} {\bibfield
  {journal} {\bibinfo  {journal} {Phys. Rev.}\ }\textbf {\bibinfo {volume}
  {46}},\ \bibinfo {pages} {1087} (\bibinfo {year} {1934})}\BibitemShut
  {NoStop}%
\bibitem [{\citenamefont {Burke}\ \emph {et~al.}(1997)\citenamefont {Burke},
  \citenamefont {Field}, \citenamefont {Horton-Smith}, \citenamefont {Spencer},
  \citenamefont {Walz}, \citenamefont {Berridge}, \citenamefont {Bugg},
  \citenamefont {Shmakov}, \citenamefont {Weidemann}, \citenamefont {Bula},
  \citenamefont {McDonald}, \citenamefont {Prebys}, \citenamefont {Bamber},
  \citenamefont {Boege}, \citenamefont {Koffas}, \citenamefont {Kotseroglou},
  \citenamefont {Melissinos}, \citenamefont {Meyerhofer}, \citenamefont
  {Reis},\ and\ \citenamefont {Ragg}}]{Burke1997}%
  \BibitemOpen
  \bibfield  {author} {\bibinfo {author} {\bibfnamefont {D.~L.}\ \bibnamefont
  {Burke}}, \bibinfo {author} {\bibfnamefont {R.~C.}\ \bibnamefont {Field}},
  \bibinfo {author} {\bibfnamefont {G.}~\bibnamefont {Horton-Smith}}, \bibinfo
  {author} {\bibfnamefont {J.~E.}\ \bibnamefont {Spencer}}, \bibinfo {author}
  {\bibfnamefont {D.}~\bibnamefont {Walz}}, \bibinfo {author} {\bibfnamefont
  {S.~C.}\ \bibnamefont {Berridge}}, \bibinfo {author} {\bibfnamefont {W.~M.}\
  \bibnamefont {Bugg}}, \bibinfo {author} {\bibfnamefont {K.}~\bibnamefont
  {Shmakov}}, \bibinfo {author} {\bibfnamefont {A.~W.}\ \bibnamefont
  {Weidemann}}, \bibinfo {author} {\bibfnamefont {C.}~\bibnamefont {Bula}},
  \bibinfo {author} {\bibfnamefont {K.~T.}\ \bibnamefont {McDonald}}, \bibinfo
  {author} {\bibfnamefont {E.~J.}\ \bibnamefont {Prebys}}, \bibinfo {author}
  {\bibfnamefont {C.}~\bibnamefont {Bamber}}, \bibinfo {author} {\bibfnamefont
  {S.~J.}\ \bibnamefont {Boege}}, \bibinfo {author} {\bibfnamefont
  {T.}~\bibnamefont {Koffas}}, \bibinfo {author} {\bibfnamefont
  {T.}~\bibnamefont {Kotseroglou}}, \bibinfo {author} {\bibfnamefont {A.~C.}\
  \bibnamefont {Melissinos}}, \bibinfo {author} {\bibfnamefont {D.~D.}\
  \bibnamefont {Meyerhofer}}, \bibinfo {author} {\bibfnamefont {D.~A.}\
  \bibnamefont {Reis}},\ and\ \bibinfo {author} {\bibfnamefont
  {W.}~\bibnamefont {Ragg}},\ }\href
  {https://doi.org/10.1103/PhysRevLett.79.1626} {\bibfield  {journal} {\bibinfo
   {journal} {Phys. Rev. Lett.}\ }\textbf {\bibinfo {volume} {79}},\ \bibinfo
  {pages} {1626} (\bibinfo {year} {1997})}\BibitemShut {NoStop}%
\bibitem [{\citenamefont {Bamber}\ \emph {et~al.}(1999)\citenamefont {Bamber},
  \citenamefont {Boege}, \citenamefont {Koffas}, \citenamefont {Kotseroglou},
  \citenamefont {Melissinos}, \citenamefont {Meyerhofer}, \citenamefont {Reis},
  \citenamefont {Ragg}, \citenamefont {Bula}, \citenamefont {McDonald},
  \citenamefont {Prebys}, \citenamefont {Burke}, \citenamefont {Field},
  \citenamefont {Horton-Smith}, \citenamefont {Spencer}, \citenamefont {Walz},
  \citenamefont {Berridge}, \citenamefont {Bugg}, \citenamefont {Shmakov},\
  and\ \citenamefont {Weidemann}}]{Bamber1999}%
  \BibitemOpen
  \bibfield  {author} {\bibinfo {author} {\bibfnamefont {C.}~\bibnamefont
  {Bamber}}, \bibinfo {author} {\bibfnamefont {S.~J.}\ \bibnamefont {Boege}},
  \bibinfo {author} {\bibfnamefont {T.}~\bibnamefont {Koffas}}, \bibinfo
  {author} {\bibfnamefont {T.}~\bibnamefont {Kotseroglou}}, \bibinfo {author}
  {\bibfnamefont {A.~C.}\ \bibnamefont {Melissinos}}, \bibinfo {author}
  {\bibfnamefont {D.~D.}\ \bibnamefont {Meyerhofer}}, \bibinfo {author}
  {\bibfnamefont {D.~A.}\ \bibnamefont {Reis}}, \bibinfo {author}
  {\bibfnamefont {W.}~\bibnamefont {Ragg}}, \bibinfo {author} {\bibfnamefont
  {C.}~\bibnamefont {Bula}}, \bibinfo {author} {\bibfnamefont {K.~T.}\
  \bibnamefont {McDonald}}, \bibinfo {author} {\bibfnamefont {E.~J.}\
  \bibnamefont {Prebys}}, \bibinfo {author} {\bibfnamefont {D.~L.}\
  \bibnamefont {Burke}}, \bibinfo {author} {\bibfnamefont {R.~C.}\ \bibnamefont
  {Field}}, \bibinfo {author} {\bibfnamefont {G.}~\bibnamefont {Horton-Smith}},
  \bibinfo {author} {\bibfnamefont {J.~E.}\ \bibnamefont {Spencer}}, \bibinfo
  {author} {\bibfnamefont {D.}~\bibnamefont {Walz}}, \bibinfo {author}
  {\bibfnamefont {S.~C.}\ \bibnamefont {Berridge}}, \bibinfo {author}
  {\bibfnamefont {W.~M.}\ \bibnamefont {Bugg}}, \bibinfo {author}
  {\bibfnamefont {K.}~\bibnamefont {Shmakov}},\ and\ \bibinfo {author}
  {\bibfnamefont {A.~W.}\ \bibnamefont {Weidemann}},\ }\href
  {https://doi.org/10.1103/PhysRevD.60.092004} {\bibfield  {journal} {\bibinfo
  {journal} {Phys. Rev. D}\ }\textbf {\bibinfo {volume} {60}},\ \bibinfo
  {pages} {092004} (\bibinfo {year} {1999})}\BibitemShut {NoStop}%
\bibitem [{\citenamefont {Blackburn}(2020)}]{Blackburn2020}%
  \BibitemOpen
  \bibfield  {author} {\bibinfo {author} {\bibfnamefont {T.~G.}\ \bibnamefont
  {Blackburn}},\ }\href {https://doi.org/10.1007/s41614-020-0042-0} {\bibfield
  {journal} {\bibinfo  {journal} {Rev. Mod. Plasma Phys.}\ }\textbf {\bibinfo
  {volume} {4}},\ \bibinfo {pages} {5} (\bibinfo {year} {2020})}\BibitemShut
  {NoStop}%
\bibitem [{\citenamefont {Gonoskov}\ \emph {et~al.}(2022)\citenamefont
  {Gonoskov}, \citenamefont {Blackburn}, \citenamefont {Marklund},\ and\
  \citenamefont {Bulanov}}]{Gonoskov2022Charged}%
  \BibitemOpen
  \bibfield  {author} {\bibinfo {author} {\bibfnamefont {A.}~\bibnamefont
  {Gonoskov}}, \bibinfo {author} {\bibfnamefont {T.~G.}\ \bibnamefont
  {Blackburn}}, \bibinfo {author} {\bibfnamefont {M.}~\bibnamefont
  {Marklund}},\ and\ \bibinfo {author} {\bibfnamefont {S.~S.}\ \bibnamefont
  {Bulanov}},\ }\href {https://doi.org/10.1103/RevModPhys.94.045001} {\bibfield
   {journal} {\bibinfo  {journal} {Rev. Mod. Phys.}\ }\textbf {\bibinfo
  {volume} {94}},\ \bibinfo {pages} {045001} (\bibinfo {year}
  {2022})}\BibitemShut {NoStop}%
\bibitem [{\citenamefont {Ruffini}\ \emph {et~al.}(2010)\citenamefont
  {Ruffini}, \citenamefont {Vereshchagin},\ and\ \citenamefont
  {Xue}}]{Ruffini2010}%
  \BibitemOpen
  \bibfield  {author} {\bibinfo {author} {\bibfnamefont {R.}~\bibnamefont
  {Ruffini}}, \bibinfo {author} {\bibfnamefont {G.}~\bibnamefont
  {Vereshchagin}},\ and\ \bibinfo {author} {\bibfnamefont {S.-S.}\ \bibnamefont
  {Xue}},\ }\href
  {https://doi.org/https://doi.org/10.1016/j.physrep.2009.10.004} {\bibfield
  {journal} {\bibinfo  {journal} {Phys. Rep.}\ }\textbf {\bibinfo {volume}
  {487}},\ \bibinfo {pages} {1} (\bibinfo {year} {2010})}\BibitemShut {NoStop}%
\bibitem [{\citenamefont {Bula}\ \emph {et~al.}(1996)\citenamefont {Bula},
  \citenamefont {McDonald}, \citenamefont {Prebys}, \citenamefont {Bamber},
  \citenamefont {Boege}, \citenamefont {Kotseroglou}, \citenamefont
  {Melissinos}, \citenamefont {Meyerhofer}, \citenamefont {Ragg}, \citenamefont
  {Burke}, \citenamefont {Field}, \citenamefont {Horton-Smith}, \citenamefont
  {Odian}, \citenamefont {Spencer}, \citenamefont {Walz}, \citenamefont
  {Berridge}, \citenamefont {Bugg}, \citenamefont {Shmakov},\ and\
  \citenamefont {Weidemann}}]{Bula1996}%
  \BibitemOpen
  \bibfield  {author} {\bibinfo {author} {\bibfnamefont {C.}~\bibnamefont
  {Bula}}, \bibinfo {author} {\bibfnamefont {K.~T.}\ \bibnamefont {McDonald}},
  \bibinfo {author} {\bibfnamefont {E.~J.}\ \bibnamefont {Prebys}}, \bibinfo
  {author} {\bibfnamefont {C.}~\bibnamefont {Bamber}}, \bibinfo {author}
  {\bibfnamefont {S.}~\bibnamefont {Boege}}, \bibinfo {author} {\bibfnamefont
  {T.}~\bibnamefont {Kotseroglou}}, \bibinfo {author} {\bibfnamefont {A.~C.}\
  \bibnamefont {Melissinos}}, \bibinfo {author} {\bibfnamefont {D.~D.}\
  \bibnamefont {Meyerhofer}}, \bibinfo {author} {\bibfnamefont
  {W.}~\bibnamefont {Ragg}}, \bibinfo {author} {\bibfnamefont {D.~L.}\
  \bibnamefont {Burke}}, \bibinfo {author} {\bibfnamefont {R.~C.}\ \bibnamefont
  {Field}}, \bibinfo {author} {\bibfnamefont {G.}~\bibnamefont {Horton-Smith}},
  \bibinfo {author} {\bibfnamefont {A.~C.}\ \bibnamefont {Odian}}, \bibinfo
  {author} {\bibfnamefont {J.~E.}\ \bibnamefont {Spencer}}, \bibinfo {author}
  {\bibfnamefont {D.}~\bibnamefont {Walz}}, \bibinfo {author} {\bibfnamefont
  {S.~C.}\ \bibnamefont {Berridge}}, \bibinfo {author} {\bibfnamefont {W.~M.}\
  \bibnamefont {Bugg}}, \bibinfo {author} {\bibfnamefont {K.}~\bibnamefont
  {Shmakov}},\ and\ \bibinfo {author} {\bibfnamefont {A.~W.}\ \bibnamefont
  {Weidemann}},\ }\href {https://doi.org/10.1103/PhysRevLett.76.3116}
  {\bibfield  {journal} {\bibinfo  {journal} {Phys. Rev. Lett.}\ }\textbf
  {\bibinfo {volume} {76}},\ \bibinfo {pages} {3116} (\bibinfo {year}
  {1996})}\BibitemShut {NoStop}%
\bibitem [{\citenamefont {Bell}\ and\ \citenamefont {Kirk}(2008)}]{Bell2008}%
  \BibitemOpen
  \bibfield  {author} {\bibinfo {author} {\bibfnamefont {A.~R.}\ \bibnamefont
  {Bell}}\ and\ \bibinfo {author} {\bibfnamefont {J.~G.}\ \bibnamefont
  {Kirk}},\ }\href {https://doi.org/10.1103/PhysRevLett.101.200403} {\bibfield
  {journal} {\bibinfo  {journal} {Phys. Rev. Lett.}\ }\textbf {\bibinfo
  {volume} {101}},\ \bibinfo {pages} {200403} (\bibinfo {year}
  {2008})}\BibitemShut {NoStop}%
\bibitem [{\citenamefont {Tamburini}\ \emph {et~al.}(2017)\citenamefont
  {Tamburini}, \citenamefont {Di~Piazza},\ and\ \citenamefont
  {Keitel}}]{Tamburini2017}%
  \BibitemOpen
  \bibfield  {author} {\bibinfo {author} {\bibfnamefont {M.}~\bibnamefont
  {Tamburini}}, \bibinfo {author} {\bibfnamefont {A.}~\bibnamefont
  {Di~Piazza}},\ and\ \bibinfo {author} {\bibfnamefont {C.~H.}\ \bibnamefont
  {Keitel}},\ }\href {https://doi.org/10.1038/s41598-017-05891-z} {\bibfield
  {journal} {\bibinfo  {journal} {Sci. Rep.}\ }\textbf {\bibinfo {volume}
  {7}},\ \bibinfo {pages} {5694} (\bibinfo {year} {2017})}\BibitemShut
  {NoStop}%
\bibitem [{\citenamefont {Danson}\ \emph {et~al.}(2019)\citenamefont {Danson},
  \citenamefont {Haefner}, \citenamefont {Bromage}, \citenamefont {Butcher},
  \citenamefont {Chanteloup}, \citenamefont {Chowdhury}, \citenamefont
  {Galvanauskas}, \citenamefont {Gizzi}, \citenamefont {Hein}, \citenamefont
  {Hillier}, \citenamefont {Hopps}, \citenamefont {Kato}, \citenamefont
  {Khazanov}, \citenamefont {Kodama}, \citenamefont {Korn}, \citenamefont {Li},
  \citenamefont {Li}, \citenamefont {Limpert}, \citenamefont {Ma},
  \citenamefont {Nam}, \citenamefont {Neely}, \citenamefont {Papadopoulos},
  \citenamefont {Penman}, \citenamefont {Qian}, \citenamefont {Rocca},
  \citenamefont {Shaykin}, \citenamefont {Siders}, \citenamefont {Spindloe},
  \citenamefont {Szatm{\'{a}}ri}, \citenamefont {Trines}, \citenamefont {Zhu},
  \citenamefont {Zhu},\ and\ \citenamefont {Zuegel}}]{Danson2019}%
  \BibitemOpen
  \bibfield  {author} {\bibinfo {author} {\bibfnamefont {C.~N.}\ \bibnamefont
  {Danson}}, \bibinfo {author} {\bibfnamefont {C.}~\bibnamefont {Haefner}},
  \bibinfo {author} {\bibfnamefont {J.}~\bibnamefont {Bromage}}, \bibinfo
  {author} {\bibfnamefont {T.}~\bibnamefont {Butcher}}, \bibinfo {author}
  {\bibfnamefont {J.-C.~F.}\ \bibnamefont {Chanteloup}}, \bibinfo {author}
  {\bibfnamefont {E.~A.}\ \bibnamefont {Chowdhury}}, \bibinfo {author}
  {\bibfnamefont {A.}~\bibnamefont {Galvanauskas}}, \bibinfo {author}
  {\bibfnamefont {L.~A.}\ \bibnamefont {Gizzi}}, \bibinfo {author}
  {\bibfnamefont {J.}~\bibnamefont {Hein}}, \bibinfo {author} {\bibfnamefont
  {D.~I.}\ \bibnamefont {Hillier}}, \bibinfo {author} {\bibfnamefont {N.~W.}\
  \bibnamefont {Hopps}}, \bibinfo {author} {\bibfnamefont {Y.}~\bibnamefont
  {Kato}}, \bibinfo {author} {\bibfnamefont {E.~A.}\ \bibnamefont {Khazanov}},
  \bibinfo {author} {\bibfnamefont {R.}~\bibnamefont {Kodama}}, \bibinfo
  {author} {\bibfnamefont {G.}~\bibnamefont {Korn}}, \bibinfo {author}
  {\bibfnamefont {R.}~\bibnamefont {Li}}, \bibinfo {author} {\bibfnamefont
  {Y.}~\bibnamefont {Li}}, \bibinfo {author} {\bibfnamefont {J.}~\bibnamefont
  {Limpert}}, \bibinfo {author} {\bibfnamefont {J.}~\bibnamefont {Ma}},
  \bibinfo {author} {\bibfnamefont {C.~H.}\ \bibnamefont {Nam}}, \bibinfo
  {author} {\bibfnamefont {D.}~\bibnamefont {Neely}}, \bibinfo {author}
  {\bibfnamefont {D.}~\bibnamefont {Papadopoulos}}, \bibinfo {author}
  {\bibfnamefont {R.~R.}\ \bibnamefont {Penman}}, \bibinfo {author}
  {\bibfnamefont {L.}~\bibnamefont {Qian}}, \bibinfo {author} {\bibfnamefont
  {J.~J.}\ \bibnamefont {Rocca}}, \bibinfo {author} {\bibfnamefont {A.~A.}\
  \bibnamefont {Shaykin}}, \bibinfo {author} {\bibfnamefont {C.~W.}\
  \bibnamefont {Siders}}, \bibinfo {author} {\bibfnamefont {C.}~\bibnamefont
  {Spindloe}}, \bibinfo {author} {\bibfnamefont {S.}~\bibnamefont
  {Szatm{\'{a}}ri}}, \bibinfo {author} {\bibfnamefont {R.~M. G.~M.}\
  \bibnamefont {Trines}}, \bibinfo {author} {\bibfnamefont {J.}~\bibnamefont
  {Zhu}}, \bibinfo {author} {\bibfnamefont {P.}~\bibnamefont {Zhu}},\ and\
  \bibinfo {author} {\bibfnamefont {J.~D.}\ \bibnamefont {Zuegel}},\ }\href
  {https://doi.org/10.1017/hpl.2019.36} {\bibfield  {journal} {\bibinfo
  {journal} {High Power Laser Sci. Eng.}\ }\textbf {\bibinfo {volume} {7}},\
  \bibinfo {pages} {e54} (\bibinfo {year} {2019})}\BibitemShut {NoStop}%
\bibitem [{\citenamefont {Yoon}\ \emph {et~al.}(2021)\citenamefont {Yoon},
  \citenamefont {Kim}, \citenamefont {Choi}, \citenamefont {Sung},
  \citenamefont {Lee}, \citenamefont {Lee},\ and\ \citenamefont
  {Nam}}]{Yoon2021}%
  \BibitemOpen
  \bibfield  {author} {\bibinfo {author} {\bibfnamefont {J.~W.}\ \bibnamefont
  {Yoon}}, \bibinfo {author} {\bibfnamefont {Y.~G.}\ \bibnamefont {Kim}},
  \bibinfo {author} {\bibfnamefont {I.~W.}\ \bibnamefont {Choi}}, \bibinfo
  {author} {\bibfnamefont {J.~H.}\ \bibnamefont {Sung}}, \bibinfo {author}
  {\bibfnamefont {H.~W.}\ \bibnamefont {Lee}}, \bibinfo {author} {\bibfnamefont
  {S.~K.}\ \bibnamefont {Lee}},\ and\ \bibinfo {author} {\bibfnamefont {C.~H.}\
  \bibnamefont {Nam}},\ }\href {https://doi.org/10.1364/OPTICA.420520}
  {\bibfield  {journal} {\bibinfo  {journal} {Optica}\ }\textbf {\bibinfo
  {volume} {8}},\ \bibinfo {pages} {630} (\bibinfo {year} {2021})}\BibitemShut
  {NoStop}%
\bibitem [{\citenamefont {Mourou}\ \emph {et~al.}(2002)\citenamefont {Mourou},
  \citenamefont {Chang}, \citenamefont {Maksimchuk}, \citenamefont {Nees},
  \citenamefont {Bulanov}, \citenamefont {Bychenkov}, \citenamefont
  {Esirkepov}, \citenamefont {Naumova}, \citenamefont {Pegoraro},\ and\
  \citenamefont {Ruhl}}]{Mourou2002}%
  \BibitemOpen
  \bibfield  {author} {\bibinfo {author} {\bibfnamefont {G.}~\bibnamefont
  {Mourou}}, \bibinfo {author} {\bibfnamefont {Z.}~\bibnamefont {Chang}},
  \bibinfo {author} {\bibfnamefont {A.}~\bibnamefont {Maksimchuk}}, \bibinfo
  {author} {\bibfnamefont {J.}~\bibnamefont {Nees}}, \bibinfo {author}
  {\bibfnamefont {S.~V.}\ \bibnamefont {Bulanov}}, \bibinfo {author}
  {\bibfnamefont {V.~Y.}\ \bibnamefont {Bychenkov}}, \bibinfo {author}
  {\bibfnamefont {T.~Z.}\ \bibnamefont {Esirkepov}}, \bibinfo {author}
  {\bibfnamefont {N.~M.}\ \bibnamefont {Naumova}}, \bibinfo {author}
  {\bibfnamefont {F.}~\bibnamefont {Pegoraro}},\ and\ \bibinfo {author}
  {\bibfnamefont {H.}~\bibnamefont {Ruhl}},\ }\href
  {https://doi.org/10.1134/1.1434292} {\bibfield  {journal} {\bibinfo
  {journal} {Plasma Phys. Rep.}\ }\textbf {\bibinfo {volume} {28}},\ \bibinfo
  {pages} {12–27} (\bibinfo {year} {2002})}\BibitemShut {NoStop}%
\bibitem [{\citenamefont {Mourou}\ \emph {et~al.}(2006)\citenamefont {Mourou},
  \citenamefont {Tajima},\ and\ \citenamefont {Bulanov}}]{Mourou2006}%
  \BibitemOpen
  \bibfield  {author} {\bibinfo {author} {\bibfnamefont {G.~A.}\ \bibnamefont
  {Mourou}}, \bibinfo {author} {\bibfnamefont {T.}~\bibnamefont {Tajima}},\
  and\ \bibinfo {author} {\bibfnamefont {S.~V.}\ \bibnamefont {Bulanov}},\
  }\href {https://doi.org/10.1103/RevModPhys.78.309} {\bibfield  {journal}
  {\bibinfo  {journal} {Rev. Mod. Phys.}\ }\textbf {\bibinfo {volume} {78}},\
  \bibinfo {pages} {309} (\bibinfo {year} {2006})}\BibitemShut {NoStop}%
\bibitem [{\citenamefont {Bulanov}\ \emph
  {et~al.}(2010{\natexlab{a}})\citenamefont {Bulanov}, \citenamefont {Mur},
  \citenamefont {Narozhny}, \citenamefont {Nees},\ and\ \citenamefont
  {Popov}}]{Bulanov2010Multiple}%
  \BibitemOpen
  \bibfield  {author} {\bibinfo {author} {\bibfnamefont {S.~S.}\ \bibnamefont
  {Bulanov}}, \bibinfo {author} {\bibfnamefont {V.~D.}\ \bibnamefont {Mur}},
  \bibinfo {author} {\bibfnamefont {N.~B.}\ \bibnamefont {Narozhny}}, \bibinfo
  {author} {\bibfnamefont {J.}~\bibnamefont {Nees}},\ and\ \bibinfo {author}
  {\bibfnamefont {V.~S.}\ \bibnamefont {Popov}},\ }\href
  {https://doi.org/10.1103/PhysRevLett.104.220404} {\bibfield  {journal}
  {\bibinfo  {journal} {Phys. Rev. Lett.}\ }\textbf {\bibinfo {volume} {104}},\
  \bibinfo {pages} {220404} (\bibinfo {year} {2010}{\natexlab{a}})}\BibitemShut
  {NoStop}%
\bibitem [{\citenamefont {Nerush}\ \emph {et~al.}(2011)\citenamefont {Nerush},
  \citenamefont {Kostyukov}, \citenamefont {Fedotov}, \citenamefont {Narozhny},
  \citenamefont {Elkina},\ and\ \citenamefont {Ruhl}}]{Nerush2011}%
  \BibitemOpen
  \bibfield  {author} {\bibinfo {author} {\bibfnamefont {E.~N.}\ \bibnamefont
  {Nerush}}, \bibinfo {author} {\bibfnamefont {I.~Y.}\ \bibnamefont
  {Kostyukov}}, \bibinfo {author} {\bibfnamefont {A.~M.}\ \bibnamefont
  {Fedotov}}, \bibinfo {author} {\bibfnamefont {N.~B.}\ \bibnamefont
  {Narozhny}}, \bibinfo {author} {\bibfnamefont {N.~V.}\ \bibnamefont
  {Elkina}},\ and\ \bibinfo {author} {\bibfnamefont {H.}~\bibnamefont {Ruhl}},\
  }\href {https://doi.org/10.1103/PhysRevLett.106.035001} {\bibfield  {journal}
  {\bibinfo  {journal} {Phys. Rev. Lett.}\ }\textbf {\bibinfo {volume} {106}},\
  \bibinfo {pages} {035001} (\bibinfo {year} {2011})}\BibitemShut {NoStop}%
\bibitem [{\citenamefont {Gelfer}\ \emph {et~al.}(2015)\citenamefont {Gelfer},
  \citenamefont {Mironov}, \citenamefont {Fedotov}, \citenamefont {Bashmakov},
  \citenamefont {Nerush}, \citenamefont {Kostyukov},\ and\ \citenamefont
  {Narozhny}}]{Gelfer2015}%
  \BibitemOpen
  \bibfield  {author} {\bibinfo {author} {\bibfnamefont {E.~G.}\ \bibnamefont
  {Gelfer}}, \bibinfo {author} {\bibfnamefont {A.~A.}\ \bibnamefont {Mironov}},
  \bibinfo {author} {\bibfnamefont {A.~M.}\ \bibnamefont {Fedotov}}, \bibinfo
  {author} {\bibfnamefont {V.~F.}\ \bibnamefont {Bashmakov}}, \bibinfo {author}
  {\bibfnamefont {E.~N.}\ \bibnamefont {Nerush}}, \bibinfo {author}
  {\bibfnamefont {I.~Y.}\ \bibnamefont {Kostyukov}},\ and\ \bibinfo {author}
  {\bibfnamefont {N.~B.}\ \bibnamefont {Narozhny}},\ }\href
  {https://doi.org/10.1103/PhysRevA.92.022113} {\bibfield  {journal} {\bibinfo
  {journal} {Phys. Rev. A}\ }\textbf {\bibinfo {volume} {92}},\ \bibinfo
  {pages} {022113} (\bibinfo {year} {2015})}\BibitemShut {NoStop}%
\bibitem [{\citenamefont {Grismayer}\ \emph {et~al.}(2016)\citenamefont
  {Grismayer}, \citenamefont {Vranic}, \citenamefont {Martins}, \citenamefont
  {Fonseca},\ and\ \citenamefont {Silva}}]{Grismayer2016}%
  \BibitemOpen
  \bibfield  {author} {\bibinfo {author} {\bibfnamefont {T.}~\bibnamefont
  {Grismayer}}, \bibinfo {author} {\bibfnamefont {M.}~\bibnamefont {Vranic}},
  \bibinfo {author} {\bibfnamefont {J.~L.}\ \bibnamefont {Martins}}, \bibinfo
  {author} {\bibfnamefont {R.~A.}\ \bibnamefont {Fonseca}},\ and\ \bibinfo
  {author} {\bibfnamefont {L.~O.}\ \bibnamefont {Silva}},\ }\href
  {https://doi.org/10.1063/1.4950841} {\bibfield  {journal} {\bibinfo
  {journal} {Phys. of Plasmas}\ }\textbf {\bibinfo {volume} {23}},\ \bibinfo
  {pages} {056706} (\bibinfo {year} {2016})}\BibitemShut {NoStop}%
\bibitem [{\citenamefont {Vranic}\ \emph {et~al.}(2016)\citenamefont {Vranic},
  \citenamefont {Grismayer}, \citenamefont {Fonseca},\ and\ \citenamefont
  {Silva}}]{Vranic2016}%
  \BibitemOpen
  \bibfield  {author} {\bibinfo {author} {\bibfnamefont {M.}~\bibnamefont
  {Vranic}}, \bibinfo {author} {\bibfnamefont {T.}~\bibnamefont {Grismayer}},
  \bibinfo {author} {\bibfnamefont {R.~A.}\ \bibnamefont {Fonseca}},\ and\
  \bibinfo {author} {\bibfnamefont {L.~O.}\ \bibnamefont {Silva}},\ }\href
  {https://doi.org/10.1088/0741-3335/59/1/014040} {\bibfield  {journal}
  {\bibinfo  {journal} {Plasma Phys. Control. Fusion}\ }\textbf {\bibinfo
  {volume} {59}},\ \bibinfo {pages} {014040} (\bibinfo {year}
  {2016})}\BibitemShut {NoStop}%
\bibitem [{\citenamefont {Gong}\ \emph {et~al.}(2017)\citenamefont {Gong},
  \citenamefont {Hu}, \citenamefont {Shou}, \citenamefont {Qiao}, \citenamefont
  {Chen}, \citenamefont {He}, \citenamefont {Bulanov}, \citenamefont
  {Esirkepov}, \citenamefont {Bulanov},\ and\ \citenamefont {Yan}}]{Gong2017}%
  \BibitemOpen
  \bibfield  {author} {\bibinfo {author} {\bibfnamefont {Z.}~\bibnamefont
  {Gong}}, \bibinfo {author} {\bibfnamefont {R.~H.}\ \bibnamefont {Hu}},
  \bibinfo {author} {\bibfnamefont {Y.~R.}\ \bibnamefont {Shou}}, \bibinfo
  {author} {\bibfnamefont {B.}~\bibnamefont {Qiao}}, \bibinfo {author}
  {\bibfnamefont {C.~E.}\ \bibnamefont {Chen}}, \bibinfo {author}
  {\bibfnamefont {X.~T.}\ \bibnamefont {He}}, \bibinfo {author} {\bibfnamefont
  {S.~S.}\ \bibnamefont {Bulanov}}, \bibinfo {author} {\bibfnamefont {T.~Z.}\
  \bibnamefont {Esirkepov}}, \bibinfo {author} {\bibfnamefont {S.~V.}\
  \bibnamefont {Bulanov}},\ and\ \bibinfo {author} {\bibfnamefont {X.~Q.}\
  \bibnamefont {Yan}},\ }\href {https://doi.org/10.1103/PhysRevE.95.013210}
  {\bibfield  {journal} {\bibinfo  {journal} {Phys. Rev. E}\ }\textbf {\bibinfo
  {volume} {95}},\ \bibinfo {pages} {013210} (\bibinfo {year}
  {2017})}\BibitemShut {NoStop}%
\bibitem [{\citenamefont {Grismayer}\ \emph {et~al.}(2017)\citenamefont
  {Grismayer}, \citenamefont {Vranic}, \citenamefont {Martins}, \citenamefont
  {Fonseca},\ and\ \citenamefont {Silva}}]{Grismayer2017}%
  \BibitemOpen
  \bibfield  {author} {\bibinfo {author} {\bibfnamefont {T.}~\bibnamefont
  {Grismayer}}, \bibinfo {author} {\bibfnamefont {M.}~\bibnamefont {Vranic}},
  \bibinfo {author} {\bibfnamefont {J.~L.}\ \bibnamefont {Martins}}, \bibinfo
  {author} {\bibfnamefont {R.~A.}\ \bibnamefont {Fonseca}},\ and\ \bibinfo
  {author} {\bibfnamefont {L.~O.}\ \bibnamefont {Silva}},\ }\href
  {https://doi.org/10.1103/PhysRevE.95.023210} {\bibfield  {journal} {\bibinfo
  {journal} {Phys. Rev. E}\ }\textbf {\bibinfo {volume} {95}},\ \bibinfo
  {pages} {023210} (\bibinfo {year} {2017})}\BibitemShut {NoStop}%
\bibitem [{\citenamefont {Jirka}\ \emph {et~al.}(2017)\citenamefont {Jirka},
  \citenamefont {Klimo}, \citenamefont {Vranic}, \citenamefont {Weber},\ and\
  \citenamefont {Korn}}]{Jirka2017}%
  \BibitemOpen
  \bibfield  {author} {\bibinfo {author} {\bibfnamefont {M.}~\bibnamefont
  {Jirka}}, \bibinfo {author} {\bibfnamefont {O.}~\bibnamefont {Klimo}},
  \bibinfo {author} {\bibfnamefont {M.}~\bibnamefont {Vranic}}, \bibinfo
  {author} {\bibfnamefont {S.}~\bibnamefont {Weber}},\ and\ \bibinfo {author}
  {\bibfnamefont {G.}~\bibnamefont {Korn}},\ }\href
  {https://doi.org/10.1038/s41598-017-15747-1} {\bibfield  {journal} {\bibinfo
  {journal} {Sci. Rep.}\ }\textbf {\bibinfo {volume} {7}},\ \bibinfo {pages}
  {15302} (\bibinfo {year} {2017})}\BibitemShut {NoStop}%
\bibitem [{\citenamefont {Fedotov}\ \emph {et~al.}(2010)\citenamefont
  {Fedotov}, \citenamefont {Narozhny}, \citenamefont {Mourou},\ and\
  \citenamefont {Korn}}]{Fedotov2010}%
  \BibitemOpen
  \bibfield  {author} {\bibinfo {author} {\bibfnamefont {A.~M.}\ \bibnamefont
  {Fedotov}}, \bibinfo {author} {\bibfnamefont {N.~B.}\ \bibnamefont
  {Narozhny}}, \bibinfo {author} {\bibfnamefont {G.}~\bibnamefont {Mourou}},\
  and\ \bibinfo {author} {\bibfnamefont {G.}~\bibnamefont {Korn}},\ }\href
  {https://doi.org/10.1103/PhysRevLett.105.080402} {\bibfield  {journal}
  {\bibinfo  {journal} {Phys. Rev. Lett.}\ }\textbf {\bibinfo {volume} {105}},\
  \bibinfo {pages} {080402} (\bibinfo {year} {2010})}\BibitemShut {NoStop}%
\bibitem [{\citenamefont {Bulanov}\ \emph
  {et~al.}(2010{\natexlab{b}})\citenamefont {Bulanov}, \citenamefont
  {Esirkepov}, \citenamefont {Thomas}, \citenamefont {Koga},\ and\
  \citenamefont {Bulanov}}]{Bulanov2010}%
  \BibitemOpen
  \bibfield  {author} {\bibinfo {author} {\bibfnamefont {S.~S.}\ \bibnamefont
  {Bulanov}}, \bibinfo {author} {\bibfnamefont {T.~Z.}\ \bibnamefont
  {Esirkepov}}, \bibinfo {author} {\bibfnamefont {A.~G.~R.}\ \bibnamefont
  {Thomas}}, \bibinfo {author} {\bibfnamefont {J.~K.}\ \bibnamefont {Koga}},\
  and\ \bibinfo {author} {\bibfnamefont {S.~V.}\ \bibnamefont {Bulanov}},\
  }\href {https://doi.org/10.1103/PhysRevLett.105.220407} {\bibfield  {journal}
  {\bibinfo  {journal} {Phys. Rev. Lett.}\ }\textbf {\bibinfo {volume} {105}},\
  \bibinfo {pages} {220407} (\bibinfo {year} {2010}{\natexlab{b}})}\BibitemShut
  {NoStop}%
\bibitem [{\citenamefont {Schwinger}(1951)}]{Schwinger1951}%
  \BibitemOpen
  \bibfield  {author} {\bibinfo {author} {\bibfnamefont {J.}~\bibnamefont
  {Schwinger}},\ }\href {https://doi.org/10.1103/PhysRev.82.664} {\bibfield
  {journal} {\bibinfo  {journal} {Phys. Rev.}\ }\textbf {\bibinfo {volume}
  {82}},\ \bibinfo {pages} {664} (\bibinfo {year} {1951})}\BibitemShut
  {NoStop}%
\bibitem [{\citenamefont {Bassett}(1986)}]{Bassett1986}%
  \BibitemOpen
  \bibfield  {author} {\bibinfo {author} {\bibfnamefont {I.}~\bibnamefont
  {Bassett}},\ }\href {https://doi.org/10.1080/713821943} {\bibfield  {journal}
  {\bibinfo  {journal} {Opt. Acta}\ }\textbf {\bibinfo {volume} {33}},\
  \bibinfo {pages} {279} (\bibinfo {year} {1986})}\BibitemShut {NoStop}%
\bibitem [{\citenamefont {Gonoskov}\ \emph {et~al.}(2012)\citenamefont
  {Gonoskov}, \citenamefont {Aiello}, \citenamefont {Heugel},\ and\
  \citenamefont {Leuchs}}]{Gonoskov2012}%
  \BibitemOpen
  \bibfield  {author} {\bibinfo {author} {\bibfnamefont {I.}~\bibnamefont
  {Gonoskov}}, \bibinfo {author} {\bibfnamefont {A.}~\bibnamefont {Aiello}},
  \bibinfo {author} {\bibfnamefont {S.}~\bibnamefont {Heugel}},\ and\ \bibinfo
  {author} {\bibfnamefont {G.}~\bibnamefont {Leuchs}},\ }\href
  {https://doi.org/10.1103/PhysRevA.86.053836} {\bibfield  {journal} {\bibinfo
  {journal} {Phys. Rev. A}\ }\textbf {\bibinfo {volume} {86}},\ \bibinfo
  {pages} {053836} (\bibinfo {year} {2012})}\BibitemShut {NoStop}%
\bibitem [{\citenamefont {Gonoskov}\ \emph {et~al.}(2013)\citenamefont
  {Gonoskov}, \citenamefont {Gonoskov}, \citenamefont {Harvey}, \citenamefont
  {Ilderton}, \citenamefont {Kim}, \citenamefont {Marklund}, \citenamefont
  {Mourou},\ and\ \citenamefont {Sergeev}}]{Gonoskov2013}%
  \BibitemOpen
  \bibfield  {author} {\bibinfo {author} {\bibfnamefont {A.}~\bibnamefont
  {Gonoskov}}, \bibinfo {author} {\bibfnamefont {I.}~\bibnamefont {Gonoskov}},
  \bibinfo {author} {\bibfnamefont {C.}~\bibnamefont {Harvey}}, \bibinfo
  {author} {\bibfnamefont {A.}~\bibnamefont {Ilderton}}, \bibinfo {author}
  {\bibfnamefont {A.}~\bibnamefont {Kim}}, \bibinfo {author} {\bibfnamefont
  {M.}~\bibnamefont {Marklund}}, \bibinfo {author} {\bibfnamefont
  {G.}~\bibnamefont {Mourou}},\ and\ \bibinfo {author} {\bibfnamefont
  {A.}~\bibnamefont {Sergeev}},\ }\href
  {https://doi.org/10.1103/PhysRevLett.111.060404} {\bibfield  {journal}
  {\bibinfo  {journal} {Phys. Rev. Lett.}\ }\textbf {\bibinfo {volume} {111}},\
  \bibinfo {pages} {060404} (\bibinfo {year} {2013})}\BibitemShut {NoStop}%
\bibitem [{\citenamefont {Gonoskov}\ \emph {et~al.}(2017)\citenamefont
  {Gonoskov}, \citenamefont {Bashinov}, \citenamefont {Bastrakov},
  \citenamefont {Efimenko}, \citenamefont {Ilderton}, \citenamefont {Kim},
  \citenamefont {Marklund}, \citenamefont {Meyerov}, \citenamefont {Muraviev},\
  and\ \citenamefont {Sergeev}}]{Gonoskov2017}%
  \BibitemOpen
  \bibfield  {author} {\bibinfo {author} {\bibfnamefont {A.}~\bibnamefont
  {Gonoskov}}, \bibinfo {author} {\bibfnamefont {A.}~\bibnamefont {Bashinov}},
  \bibinfo {author} {\bibfnamefont {S.}~\bibnamefont {Bastrakov}}, \bibinfo
  {author} {\bibfnamefont {E.}~\bibnamefont {Efimenko}}, \bibinfo {author}
  {\bibfnamefont {A.}~\bibnamefont {Ilderton}}, \bibinfo {author}
  {\bibfnamefont {A.}~\bibnamefont {Kim}}, \bibinfo {author} {\bibfnamefont
  {M.}~\bibnamefont {Marklund}}, \bibinfo {author} {\bibfnamefont
  {I.}~\bibnamefont {Meyerov}}, \bibinfo {author} {\bibfnamefont
  {A.}~\bibnamefont {Muraviev}},\ and\ \bibinfo {author} {\bibfnamefont
  {A.}~\bibnamefont {Sergeev}},\ }\href
  {https://doi.org/10.1103/PhysRevX.7.041003} {\bibfield  {journal} {\bibinfo
  {journal} {Phys. Rev. X}\ }\textbf {\bibinfo {volume} {7}},\ \bibinfo {pages}
  {041003} (\bibinfo {year} {2017})}\BibitemShut {NoStop}%
\bibitem [{\citenamefont {Efimenko}\ \emph {et~al.}(2018)\citenamefont
  {Efimenko}, \citenamefont {Bashinov}, \citenamefont {Bastrakov},
  \citenamefont {Gonoskov}, \citenamefont {Muraviev}, \citenamefont {Meyerov},
  \citenamefont {Kim},\ and\ \citenamefont {Sergeev}}]{Efimenko2018}%
  \BibitemOpen
  \bibfield  {author} {\bibinfo {author} {\bibfnamefont {E.~S.}\ \bibnamefont
  {Efimenko}}, \bibinfo {author} {\bibfnamefont {A.~V.}\ \bibnamefont
  {Bashinov}}, \bibinfo {author} {\bibfnamefont {S.~I.}\ \bibnamefont
  {Bastrakov}}, \bibinfo {author} {\bibfnamefont {A.~A.}\ \bibnamefont
  {Gonoskov}}, \bibinfo {author} {\bibfnamefont {A.~A.}\ \bibnamefont
  {Muraviev}}, \bibinfo {author} {\bibfnamefont {I.~B.}\ \bibnamefont
  {Meyerov}}, \bibinfo {author} {\bibfnamefont {A.~V.}\ \bibnamefont {Kim}},\
  and\ \bibinfo {author} {\bibfnamefont {A.~M.}\ \bibnamefont {Sergeev}},\
  }\href {https://doi.org/10.1038/s41598-018-20745-y} {\bibfield  {journal}
  {\bibinfo  {journal} {Sci. Rep.}\ }\textbf {\bibinfo {volume} {8}},\ \bibinfo
  {pages} {2329} (\bibinfo {year} {2018})}\BibitemShut {NoStop}%
\bibitem [{\citenamefont {Magnusson}\ \emph {et~al.}(2019)\citenamefont
  {Magnusson}, \citenamefont {Gonoskov}, \citenamefont {Marklund},
  \citenamefont {Esirkepov}, \citenamefont {Koga}, \citenamefont {Kondo},
  \citenamefont {Kando}, \citenamefont {Bulanov}, \citenamefont {Korn},\ and\
  \citenamefont {Bulanov}}]{Magnusson2019}%
  \BibitemOpen
  \bibfield  {author} {\bibinfo {author} {\bibfnamefont {J.}~\bibnamefont
  {Magnusson}}, \bibinfo {author} {\bibfnamefont {A.}~\bibnamefont {Gonoskov}},
  \bibinfo {author} {\bibfnamefont {M.}~\bibnamefont {Marklund}}, \bibinfo
  {author} {\bibfnamefont {T.~Z.}\ \bibnamefont {Esirkepov}}, \bibinfo {author}
  {\bibfnamefont {J.~K.}\ \bibnamefont {Koga}}, \bibinfo {author}
  {\bibfnamefont {K.}~\bibnamefont {Kondo}}, \bibinfo {author} {\bibfnamefont
  {M.}~\bibnamefont {Kando}}, \bibinfo {author} {\bibfnamefont {S.~V.}\
  \bibnamefont {Bulanov}}, \bibinfo {author} {\bibfnamefont {G.}~\bibnamefont
  {Korn}},\ and\ \bibinfo {author} {\bibfnamefont {S.~S.}\ \bibnamefont
  {Bulanov}},\ }\href {https://doi.org/10.1103/PhysRevLett.122.254801}
  {\bibfield  {journal} {\bibinfo  {journal} {Phys. Rev. Lett.}\ }\textbf
  {\bibinfo {volume} {122}},\ \bibinfo {pages} {254801} (\bibinfo {year}
  {2019})}\BibitemShut {NoStop}%
\bibitem [{\citenamefont {Efimenko}\ \emph {et~al.}(2022)\citenamefont
  {Efimenko}, \citenamefont {Bashinov}, \citenamefont {Muraviev}, \citenamefont
  {Volokitin}, \citenamefont {Meyerov}, \citenamefont {Leuchs}, \citenamefont
  {Sergeev},\ and\ \citenamefont {Kim}}]{Efimenko2022}%
  \BibitemOpen
  \bibfield  {author} {\bibinfo {author} {\bibfnamefont {E.~S.}\ \bibnamefont
  {Efimenko}}, \bibinfo {author} {\bibfnamefont {A.~V.}\ \bibnamefont
  {Bashinov}}, \bibinfo {author} {\bibfnamefont {A.~A.}\ \bibnamefont
  {Muraviev}}, \bibinfo {author} {\bibfnamefont {V.~D.}\ \bibnamefont
  {Volokitin}}, \bibinfo {author} {\bibfnamefont {I.~B.}\ \bibnamefont
  {Meyerov}}, \bibinfo {author} {\bibfnamefont {G.}~\bibnamefont {Leuchs}},
  \bibinfo {author} {\bibfnamefont {A.~M.}\ \bibnamefont {Sergeev}},\ and\
  \bibinfo {author} {\bibfnamefont {A.~V.}\ \bibnamefont {Kim}},\ }\href
  {https://doi.org/10.1103/PhysRevE.106.015201} {\bibfield  {journal} {\bibinfo
   {journal} {Phys. Rev. E}\ }\textbf {\bibinfo {volume} {106}},\ \bibinfo
  {pages} {015201} (\bibinfo {year} {2022})}\BibitemShut {NoStop}%
\bibitem [{\citenamefont {Marklund}\ \emph {et~al.}(2023)\citenamefont
  {Marklund}, \citenamefont {Blackburn}, \citenamefont {Gonoskov},
  \citenamefont {Magnusson}, \citenamefont {Bulanov},\ and\ \citenamefont
  {Ilderton}}]{Marklund2023}%
  \BibitemOpen
  \bibfield  {author} {\bibinfo {author} {\bibfnamefont {M.}~\bibnamefont
  {Marklund}}, \bibinfo {author} {\bibfnamefont {T.~G.}\ \bibnamefont
  {Blackburn}}, \bibinfo {author} {\bibfnamefont {A.}~\bibnamefont {Gonoskov}},
  \bibinfo {author} {\bibfnamefont {J.}~\bibnamefont {Magnusson}}, \bibinfo
  {author} {\bibfnamefont {S.~S.}\ \bibnamefont {Bulanov}},\ and\ \bibinfo
  {author} {\bibfnamefont {A.}~\bibnamefont {Ilderton}},\ }\href
  {https://doi.org/10.1017/hpl.2022.46} {\bibfield  {journal} {\bibinfo
  {journal} {High Power Laser Sci. Eng.}\ }\textbf {\bibinfo {volume} {11}},\
  \bibinfo {pages} {e19} (\bibinfo {year} {2023})}\BibitemShut {NoStop}%
\bibitem [{\citenamefont {Efimenko}\ \emph {et~al.}(2019)\citenamefont
  {Efimenko}, \citenamefont {Bashinov}, \citenamefont {Gonoskov}, \citenamefont
  {Bastrakov}, \citenamefont {Muraviev}, \citenamefont {Meyerov}, \citenamefont
  {Kim},\ and\ \citenamefont {Sergeev}}]{Efimenko2019}%
  \BibitemOpen
  \bibfield  {author} {\bibinfo {author} {\bibfnamefont {E.~S.}\ \bibnamefont
  {Efimenko}}, \bibinfo {author} {\bibfnamefont {A.~V.}\ \bibnamefont
  {Bashinov}}, \bibinfo {author} {\bibfnamefont {A.~A.}\ \bibnamefont
  {Gonoskov}}, \bibinfo {author} {\bibfnamefont {S.~I.}\ \bibnamefont
  {Bastrakov}}, \bibinfo {author} {\bibfnamefont {A.~A.}\ \bibnamefont
  {Muraviev}}, \bibinfo {author} {\bibfnamefont {I.~B.}\ \bibnamefont
  {Meyerov}}, \bibinfo {author} {\bibfnamefont {A.~V.}\ \bibnamefont {Kim}},\
  and\ \bibinfo {author} {\bibfnamefont {A.~M.}\ \bibnamefont {Sergeev}},\
  }\href {https://doi.org/10.1103/PhysRevE.99.031201} {\bibfield  {journal}
  {\bibinfo  {journal} {Phys. Rev. E}\ }\textbf {\bibinfo {volume} {99}},\
  \bibinfo {pages} {031201} (\bibinfo {year} {2019})}\BibitemShut {NoStop}%
\bibitem [{\citenamefont {Sampath}\ and\ \citenamefont
  {Tamburini}(2018)}]{Sampath2018}%
  \BibitemOpen
  \bibfield  {author} {\bibinfo {author} {\bibfnamefont {A.}~\bibnamefont
  {Sampath}}\ and\ \bibinfo {author} {\bibfnamefont {M.}~\bibnamefont
  {Tamburini}},\ }\href {https://doi.org/10.1063/1.5022640} {\bibfield
  {journal} {\bibinfo  {journal} {Phys. Plasmas}\ }\textbf {\bibinfo {volume}
  {25}},\ \bibinfo {pages} {083104} (\bibinfo {year} {2018})}\BibitemShut
  {NoStop}%
\bibitem [{\citenamefont {Lindlein}\ \emph {et~al.}(2007)\citenamefont
  {Lindlein}, \citenamefont {Maiwald}, \citenamefont {Konermann}, \citenamefont
  {Sondermann}, \citenamefont {Peschel},\ and\ \citenamefont
  {Leuchs}}]{Lindlein2007}%
  \BibitemOpen
  \bibfield  {author} {\bibinfo {author} {\bibfnamefont {N.}~\bibnamefont
  {Lindlein}}, \bibinfo {author} {\bibfnamefont {R.}~\bibnamefont {Maiwald}},
  \bibinfo {author} {\bibfnamefont {H.}~\bibnamefont {Konermann}}, \bibinfo
  {author} {\bibfnamefont {M.}~\bibnamefont {Sondermann}}, \bibinfo {author}
  {\bibfnamefont {U.}~\bibnamefont {Peschel}},\ and\ \bibinfo {author}
  {\bibfnamefont {G.}~\bibnamefont {Leuchs}},\ }\href
  {https://doi.org/10.1134/s1054660x07070055} {\bibfield  {journal} {\bibinfo
  {journal} {Laser Phys.}\ }\textbf {\bibinfo {volume} {17}},\ \bibinfo {pages}
  {927–934} (\bibinfo {year} {2007})}\BibitemShut {NoStop}%
\bibitem [{\citenamefont {Bashinov}\ \emph {et~al.}(2022)\citenamefont
  {Bashinov}, \citenamefont {Efimenko}, \citenamefont {Muraviev}, \citenamefont
  {Volokitin}, \citenamefont {Meyerov}, \citenamefont {Leuchs}, \citenamefont
  {Sergeev},\ and\ \citenamefont {Kim}}]{Bashinov2022}%
  \BibitemOpen
  \bibfield  {author} {\bibinfo {author} {\bibfnamefont {A.~V.}\ \bibnamefont
  {Bashinov}}, \bibinfo {author} {\bibfnamefont {E.~S.}\ \bibnamefont
  {Efimenko}}, \bibinfo {author} {\bibfnamefont {A.~A.}\ \bibnamefont
  {Muraviev}}, \bibinfo {author} {\bibfnamefont {V.~D.}\ \bibnamefont
  {Volokitin}}, \bibinfo {author} {\bibfnamefont {I.~B.}\ \bibnamefont
  {Meyerov}}, \bibinfo {author} {\bibfnamefont {G.}~\bibnamefont {Leuchs}},
  \bibinfo {author} {\bibfnamefont {A.~M.}\ \bibnamefont {Sergeev}},\ and\
  \bibinfo {author} {\bibfnamefont {A.~V.}\ \bibnamefont {Kim}},\ }\href
  {https://doi.org/10.1103/PhysRevE.105.065202} {\bibfield  {journal} {\bibinfo
   {journal} {Phys. Rev. E}\ }\textbf {\bibinfo {volume} {105}},\ \bibinfo
  {pages} {065202} (\bibinfo {year} {2022})}\BibitemShut {NoStop}%
\bibitem [{\citenamefont {Salamin}\ and\ \citenamefont
  {Keitel}(2002)}]{Salamin2002}%
  \BibitemOpen
  \bibfield  {author} {\bibinfo {author} {\bibfnamefont {Y.~I.}\ \bibnamefont
  {Salamin}}\ and\ \bibinfo {author} {\bibfnamefont {C.~H.}\ \bibnamefont
  {Keitel}},\ }\href {https://doi.org/10.1103/PhysRevLett.88.095005} {\bibfield
   {journal} {\bibinfo  {journal} {Phys. Rev. Lett.}\ }\textbf {\bibinfo
  {volume} {88}},\ \bibinfo {pages} {095005} (\bibinfo {year}
  {2002})}\BibitemShut {NoStop}%
\bibitem [{\citenamefont {Salamin}(2006{\natexlab{a}})}]{Salamin2006}%
  \BibitemOpen
  \bibfield  {author} {\bibinfo {author} {\bibfnamefont {Y.~I.}\ \bibnamefont
  {Salamin}},\ }\href {https://doi.org/10.1088/1367-2630/8/8/133} {\bibfield
  {journal} {\bibinfo  {journal} {New J. Phys.}\ }\textbf {\bibinfo {volume}
  {8}},\ \bibinfo {pages} {133–133} (\bibinfo {year}
  {2006}{\natexlab{a}})}\BibitemShut {NoStop}%
\bibitem [{\citenamefont {Salamin}(2006{\natexlab{b}})}]{Salamin2006Electron}%
  \BibitemOpen
  \bibfield  {author} {\bibinfo {author} {\bibfnamefont {Y.~I.}\ \bibnamefont
  {Salamin}},\ }\href {https://doi.org/10.1103/PhysRevA.73.043402} {\bibfield
  {journal} {\bibinfo  {journal} {Phys. Rev. A}\ }\textbf {\bibinfo {volume}
  {73}},\ \bibinfo {pages} {043402} (\bibinfo {year}
  {2006}{\natexlab{b}})}\BibitemShut {NoStop}%
\bibitem [{\citenamefont {Jirka}\ \emph {et~al.}(2022)\citenamefont {Jirka},
  \citenamefont {Sasorov},\ and\ \citenamefont {Bulanov}}]{Jirka2022}%
  \BibitemOpen
  \bibfield  {author} {\bibinfo {author} {\bibfnamefont {M.}~\bibnamefont
  {Jirka}}, \bibinfo {author} {\bibfnamefont {P.}~\bibnamefont {Sasorov}},\
  and\ \bibinfo {author} {\bibfnamefont {S.~V.}\ \bibnamefont {Bulanov}},\
  }\href {https://doi.org/10.1103/PhysRevD.105.113004} {\bibfield  {journal}
  {\bibinfo  {journal} {Phys. Rev. D}\ }\textbf {\bibinfo {volume} {105}},\
  \bibinfo {pages} {113004} (\bibinfo {year} {2022})}\BibitemShut {NoStop}%
\bibitem [{\citenamefont {Jirka}\ and\ \citenamefont
  {Kadlecová}(2023)}]{Jirka2023Pair}%
  \BibitemOpen
  \bibfield  {author} {\bibinfo {author} {\bibfnamefont {M.}~\bibnamefont
  {Jirka}}\ and\ \bibinfo {author} {\bibfnamefont {H.}~\bibnamefont
  {Kadlecová}},\ }\href {https://doi.org/10.1063/5.0168022} {\bibfield
  {journal} {\bibinfo  {journal} {Phys. Plasmas}\ }\textbf {\bibinfo {volume}
  {30}},\ \bibinfo {pages} {113102} (\bibinfo {year} {2023})}\BibitemShut
  {NoStop}%
\bibitem [{\citenamefont {Salamin}(2007)}]{Salamin2007}%
  \BibitemOpen
  \bibfield  {author} {\bibinfo {author} {\bibfnamefont {Y.}~\bibnamefont
  {Salamin}},\ }\href {https://doi.org/10.1007/s00340-006-2442-4} {\bibfield
  {journal} {\bibinfo  {journal} {Appl. Phys. B}\ }\textbf {\bibinfo {volume}
  {86}},\ \bibinfo {pages} {319–326} (\bibinfo {year} {2007})}\BibitemShut
  {NoStop}%
\bibitem [{\citenamefont {Ritus}(1985)}]{Ritus1985}%
  \BibitemOpen
  \bibfield  {author} {\bibinfo {author} {\bibfnamefont {V.~I.}\ \bibnamefont
  {Ritus}},\ }\href {https://doi.org/10.1007/bf01120220} {\bibfield  {journal}
  {\bibinfo  {journal} {J. Russ. Laser. Res.}\ }\textbf {\bibinfo {volume}
  {6}},\ \bibinfo {pages} {497} (\bibinfo {year} {1985})}\BibitemShut {NoStop}%
\bibitem [{\citenamefont {Derouillat}\ \emph {et~al.}(2018)\citenamefont
  {Derouillat}, \citenamefont {Beck}, \citenamefont {P{\'{e}}rez},
  \citenamefont {Vinci}, \citenamefont {Chiaramello}, \citenamefont {Grassi},
  \citenamefont {Fl{\'{e}}}, \citenamefont {Bouchard}, \citenamefont
  {Plotnikov}, \citenamefont {Aunai}, \citenamefont {Dargent}, \citenamefont
  {Riconda},\ and\ \citenamefont {Grech}}]{Derouillat2018}%
  \BibitemOpen
  \bibfield  {author} {\bibinfo {author} {\bibfnamefont {J.}~\bibnamefont
  {Derouillat}}, \bibinfo {author} {\bibfnamefont {A.}~\bibnamefont {Beck}},
  \bibinfo {author} {\bibfnamefont {F.}~\bibnamefont {P{\'{e}}rez}}, \bibinfo
  {author} {\bibfnamefont {T.}~\bibnamefont {Vinci}}, \bibinfo {author}
  {\bibfnamefont {M.}~\bibnamefont {Chiaramello}}, \bibinfo {author}
  {\bibfnamefont {A.}~\bibnamefont {Grassi}}, \bibinfo {author} {\bibfnamefont
  {M.}~\bibnamefont {Fl{\'{e}}}}, \bibinfo {author} {\bibfnamefont
  {G.}~\bibnamefont {Bouchard}}, \bibinfo {author} {\bibfnamefont
  {I.}~\bibnamefont {Plotnikov}}, \bibinfo {author} {\bibfnamefont
  {N.}~\bibnamefont {Aunai}}, \bibinfo {author} {\bibfnamefont
  {J.}~\bibnamefont {Dargent}}, \bibinfo {author} {\bibfnamefont
  {C.}~\bibnamefont {Riconda}},\ and\ \bibinfo {author} {\bibfnamefont
  {M.}~\bibnamefont {Grech}},\ }\href
  {https://doi.org/10.1016/j.cpc.2017.09.024} {\bibfield  {journal} {\bibinfo
  {journal} {Comput. Phys. Commun.}\ }\textbf {\bibinfo {volume} {222}},\
  \bibinfo {pages} {351} (\bibinfo {year} {2018})}\BibitemShut {NoStop}%
\bibitem [{sup()}]{supp}%
  \BibitemOpen
  \href@noop {} {}\bibinfo {note} {See Supplemental Material at
  URL-will-be-inserted-by-publisher for derivation intensity threshold and of
  Eqs. (6) and (7).}\BibitemShut {Stop}%
\bibitem [{\citenamefont {Bashmakov}\ \emph {et~al.}(2014)\citenamefont
  {Bashmakov}, \citenamefont {Nerush}, \citenamefont {Kostyukov}, \citenamefont
  {Fedotov},\ and\ \citenamefont {Narozhny}}]{Bashmakov2014}%
  \BibitemOpen
  \bibfield  {author} {\bibinfo {author} {\bibfnamefont {V.~F.}\ \bibnamefont
  {Bashmakov}}, \bibinfo {author} {\bibfnamefont {E.~N.}\ \bibnamefont
  {Nerush}}, \bibinfo {author} {\bibfnamefont {I.~Y.}\ \bibnamefont
  {Kostyukov}}, \bibinfo {author} {\bibfnamefont {A.~M.}\ \bibnamefont
  {Fedotov}},\ and\ \bibinfo {author} {\bibfnamefont {N.~B.}\ \bibnamefont
  {Narozhny}},\ }\href {https://doi.org/10.1063/1.4861863} {\bibfield
  {journal} {\bibinfo  {journal} {Phys. Plasmas}\ }\textbf {\bibinfo {volume}
  {21}},\ \bibinfo {pages} {013105} (\bibinfo {year} {2014})}\BibitemShut
  {NoStop}%
\bibitem [{\citenamefont {Mironov}\ \emph {et~al.}(2021)\citenamefont
  {Mironov}, \citenamefont {Gelfer},\ and\ \citenamefont
  {Fedotov}}]{Mironov2021}%
  \BibitemOpen
  \bibfield  {author} {\bibinfo {author} {\bibfnamefont {A.~A.}\ \bibnamefont
  {Mironov}}, \bibinfo {author} {\bibfnamefont {E.~G.}\ \bibnamefont
  {Gelfer}},\ and\ \bibinfo {author} {\bibfnamefont {A.~M.}\ \bibnamefont
  {Fedotov}},\ }\href {https://doi.org/10.1103/PhysRevA.104.012221} {\bibfield
  {journal} {\bibinfo  {journal} {Phys. Rev. A}\ }\textbf {\bibinfo {volume}
  {104}},\ \bibinfo {pages} {012221} (\bibinfo {year} {2021})}\BibitemShut
  {NoStop}%
\bibitem [{\citenamefont {Salamin}\ \emph {et~al.}(2008)\citenamefont
  {Salamin}, \citenamefont {Harman},\ and\ \citenamefont
  {Keitel}}]{Salamin2008}%
  \BibitemOpen
  \bibfield  {author} {\bibinfo {author} {\bibfnamefont {Y.~I.}\ \bibnamefont
  {Salamin}}, \bibinfo {author} {\bibfnamefont {Z.}~\bibnamefont {Harman}},\
  and\ \bibinfo {author} {\bibfnamefont {C.~H.}\ \bibnamefont {Keitel}},\
  }\href {https://doi.org/10.1103/PhysRevLett.100.155004} {\bibfield  {journal}
  {\bibinfo  {journal} {Phys. Rev. Lett.}\ }\textbf {\bibinfo {volume} {100}},\
  \bibinfo {pages} {155004} (\bibinfo {year} {2008})}\BibitemShut {NoStop}%
\bibitem [{\citenamefont {Carbajo}\ \emph {et~al.}(2014)\citenamefont
  {Carbajo}, \citenamefont {Granados}, \citenamefont {Schimpf}, \citenamefont
  {Sell}, \citenamefont {Hong}, \citenamefont {Moses},\ and\ \citenamefont
  {K\"{a}rtner}}]{Carbajo2014}%
  \BibitemOpen
  \bibfield  {author} {\bibinfo {author} {\bibfnamefont {S.}~\bibnamefont
  {Carbajo}}, \bibinfo {author} {\bibfnamefont {E.}~\bibnamefont {Granados}},
  \bibinfo {author} {\bibfnamefont {D.}~\bibnamefont {Schimpf}}, \bibinfo
  {author} {\bibfnamefont {A.}~\bibnamefont {Sell}}, \bibinfo {author}
  {\bibfnamefont {K.-H.}\ \bibnamefont {Hong}}, \bibinfo {author}
  {\bibfnamefont {J.}~\bibnamefont {Moses}},\ and\ \bibinfo {author}
  {\bibfnamefont {F.~X.}\ \bibnamefont {K\"{a}rtner}},\ }\href
  {https://doi.org/10.1364/OL.39.002487} {\bibfield  {journal} {\bibinfo
  {journal} {Opt. Lett.}\ }\textbf {\bibinfo {volume} {39}},\ \bibinfo {pages}
  {2487} (\bibinfo {year} {2014})}\BibitemShut {NoStop}%
\bibitem [{\citenamefont {Leblanc}\ \emph {et~al.}(2017)\citenamefont
  {Leblanc}, \citenamefont {Denoeud}, \citenamefont {Chopineau}, \citenamefont
  {Mennerat}, \citenamefont {Martin},\ and\ \citenamefont
  {Quéré}}]{Leblanc2017}%
  \BibitemOpen
  \bibfield  {author} {\bibinfo {author} {\bibfnamefont {A.}~\bibnamefont
  {Leblanc}}, \bibinfo {author} {\bibfnamefont {A.}~\bibnamefont {Denoeud}},
  \bibinfo {author} {\bibfnamefont {L.}~\bibnamefont {Chopineau}}, \bibinfo
  {author} {\bibfnamefont {G.}~\bibnamefont {Mennerat}}, \bibinfo {author}
  {\bibfnamefont {P.}~\bibnamefont {Martin}},\ and\ \bibinfo {author}
  {\bibfnamefont {F.}~\bibnamefont {Quéré}},\ }\href
  {https://doi.org/10.1038/nphys4007} {\bibfield  {journal} {\bibinfo
  {journal} {Nat. Phys.}\ }\textbf {\bibinfo {volume} {13}},\ \bibinfo {pages}
  {440–443} (\bibinfo {year} {2017})}\BibitemShut {NoStop}%
\bibitem [{\citenamefont {Porat}\ \emph {et~al.}(2022)\citenamefont {Porat},
  \citenamefont {Lightman}, \citenamefont {Cohen},\ and\ \citenamefont
  {Pomerantz}}]{Porat2022}%
  \BibitemOpen
  \bibfield  {author} {\bibinfo {author} {\bibfnamefont {E.}~\bibnamefont
  {Porat}}, \bibinfo {author} {\bibfnamefont {S.}~\bibnamefont {Lightman}},
  \bibinfo {author} {\bibfnamefont {I.}~\bibnamefont {Cohen}},\ and\ \bibinfo
  {author} {\bibfnamefont {I.}~\bibnamefont {Pomerantz}},\ }\href
  {https://doi.org/10.1088/2040-8986/ac79ba} {\bibfield  {journal} {\bibinfo
  {journal} {J.Opt.}\ }\textbf {\bibinfo {volume} {24}},\ \bibinfo {pages}
  {085501} (\bibinfo {year} {2022})}\BibitemShut {NoStop}%
\bibitem [{\citenamefont {Guo}\ \emph {et~al.}(2023)\citenamefont {Guo},
  \citenamefont {Zhang}, \citenamefont {Xu}, \citenamefont {Chen},
  \citenamefont {Guo}, \citenamefont {Lan},\ and\ \citenamefont
  {Shen}}]{Guo2023}%
  \BibitemOpen
  \bibfield  {author} {\bibinfo {author} {\bibfnamefont {X.}~\bibnamefont
  {Guo}}, \bibinfo {author} {\bibfnamefont {X.}~\bibnamefont {Zhang}}, \bibinfo
  {author} {\bibfnamefont {D.}~\bibnamefont {Xu}}, \bibinfo {author}
  {\bibfnamefont {W.}~\bibnamefont {Chen}}, \bibinfo {author} {\bibfnamefont
  {Y.}~\bibnamefont {Guo}}, \bibinfo {author} {\bibfnamefont {K.}~\bibnamefont
  {Lan}},\ and\ \bibinfo {author} {\bibfnamefont {B.}~\bibnamefont {Shen}},\
  }\href {https://doi.org/10.1038/s41598-023-28216-9} {\bibfield  {journal}
  {\bibinfo  {journal} {Sci. Rep.}\ }\textbf {\bibinfo {volume} {13}},\
  \bibinfo {pages} {1104} (\bibinfo {year} {2023})}\BibitemShut {NoStop}%
\bibitem [{\citenamefont {Payeur}\ \emph {et~al.}(2012)\citenamefont {Payeur},
  \citenamefont {Fourmaux}, \citenamefont {Schmidt}, \citenamefont {MacLean},
  \citenamefont {Tchervenkov}, \citenamefont {Légaré}, \citenamefont
  {Piché},\ and\ \citenamefont {Kieffer}}]{Payeur2012}%
  \BibitemOpen
  \bibfield  {author} {\bibinfo {author} {\bibfnamefont {S.}~\bibnamefont
  {Payeur}}, \bibinfo {author} {\bibfnamefont {S.}~\bibnamefont {Fourmaux}},
  \bibinfo {author} {\bibfnamefont {B.~E.}\ \bibnamefont {Schmidt}}, \bibinfo
  {author} {\bibfnamefont {J.~P.}\ \bibnamefont {MacLean}}, \bibinfo {author}
  {\bibfnamefont {C.}~\bibnamefont {Tchervenkov}}, \bibinfo {author}
  {\bibfnamefont {F.}~\bibnamefont {Légaré}}, \bibinfo {author}
  {\bibfnamefont {M.}~\bibnamefont {Piché}},\ and\ \bibinfo {author}
  {\bibfnamefont {J.~C.}\ \bibnamefont {Kieffer}},\ }\href
  {https://doi.org/10.1063/1.4738998} {\bibfield  {journal} {\bibinfo
  {journal} {Appl. Phys. Lett.}\ }\textbf {\bibinfo {volume} {101}},\ \bibinfo
  {pages} {041105} (\bibinfo {year} {2012})}\BibitemShut {NoStop}%
\end{thebibliography}%
\bibliographystyle{apsrev4-2}
\end{document}